\documentclass[natbib,letterpaper,iop]{emulateapj}
\usepackage{amstext}
\usepackage{apjfonts}
\usepackage{amsmath}
\usepackage{amssymb}
\usepackage[colorlinks=true,linkcolor=blue,citecolor=blue]{hyperref}
\begin{document}

\newcommand{\beq}{\begin{equation}}
\newcommand{\eeq}{\end{equation}}
\newcommand{\beqar}{\begin{eqnarray}}
\newcommand{\eeqar}{\end{eqnarray}}
\newcommand{\e}{\varepsilon}
\newcommand{\rt}{r_{\rm t}}
\newcommand{\rs}{r_{\rm s}}
\newcommand{\Mbh}{M_{\rm bh}}
\newcommand{\rp}{r_{\rm p}}
\newcommand{\risco}{r_{\rm isco}}
\newcommand{\rb}{r_{\rm b}}
\newcommand{\tdyn}{t_{\rm dyn}}
\newcommand{\tfb}{\tau_{\rm fb}}
\newcommand{\tpeak}{\tau_{\rm peak}}
\newcommand{\Jc}{J_{\rm c}}
\newcommand{\Jlc}{J_{\rm lc}}
\newcommand{\tr}{t_{\rm r}}
\newcommand{\tJ}{\tau_{\rm J}}
\newcommand{\Flc}{F_{\rm lc}}
\newcommand{\dmA}{\Delta m_{\rm A}}
\newcommand{\dmB}{\Delta m_{\rm B}}
\newcommand{\Ledd}{L_{\rm Edd}}
\newcommand{\torb}{\tau_{\rm orb}}
\newcommand{\tevol}{\tau_{\rm evol}}
\newcommand{\tcoll}{\tau_{\rm coll}}
\newcommand{\te}{\tau_{\e}}

\title{Illuminating Massive Black Holes With White Dwarfs:  Orbital Dynamics and High Energy Transients  from Tidal Interactions} 

\author{Morgan MacLeod$^{1}$, Jacqueline Goldstein$^{1}$, Enrico Ramirez-Ruiz$^{1}$, James Guillochon$^{2,3}$, and Johan Samsing$^{4}$}
\altaffiltext{1}{Department of Astronomy and
  Astrophysics, University of California, Santa Cruz, CA
  95064}
\altaffiltext{2}{Astronomy Department, Harvard University, 60 Garden St, Cambridge, MA 02138, USA}
\altaffiltext{3}{Einstein Fellow}
\altaffiltext{4}{Dark Cosmology Centre, Niels Bohr Institute, University of Copenhagen, 
Juliane Maries Vej 30, 2100 Copenhagen, Denmark}
   
\begin{abstract} 
White dwarfs (WDs) can be tidally disrupted only by massive black holes (MBHs) with masses less than $\sim10^5 M_\odot$. These tidal interactions feed material to the MBH well above its Eddington limit, with the potential to launch a relativistic jet. The corresponding beamed emission is a promising signpost to an otherwise quiescent MBH of relatively low mass.  
We show that the mass transfer history, and thus the lightcurve, are quite different when the disruptive orbit is parabolic, eccentric, or circular. 
The mass lost each orbit exponentiates in the eccentric-orbit case leading to the destruction of the WD after several tens of orbits.
We examine the stellar dynamics of clusters surrounding MBHs to show that single-passage WD disruptions are substantially more common than repeating encounters. 
The $10^{49}$ erg s$^{-1}$ peak luminosity of these events makes them visible to cosmological distances. 
They may be detectible at rates of as many as tens per year by instruments like {\it Swift}. 
In fact, WD-disruption transients significantly outshine their main-sequence star counterparts, and are the most likely tidal interaction to be detected arising from MBHs with masses less than $10^5 M_\odot$. 
The detection or non-detection of such WD-disruption transients by {\it Swift} is, therefore, a powerful tool to constrain lower end of the MBH mass function. 
The emerging class of ultra-long gamma ray bursts all have peak luminosities and durations reminiscent of WD disruptions, offering a hint that WD-disruption transients may already be present in existing datasets.
\end{abstract}

\maketitle

\section{Introduction}
Tidal disruption events have been  studied theoretically since their prediction by \citet{Hills:1975kh}. They are expected to produce luminous but short-lived accretion flares as debris from the disrupted star streams back to the massive black hole (MBH) \citep{Rees1988}. 
The characteristic pericenter distance for tidal disruption to occur is the tidal radius, $\rt = (\Mbh/M_*)^{1/3} R_*$, which is defined by the average density of the star and by the MBH mass.  The tidal radius scales differently with MBH mass than the MBH's Schwarzschild radius, $\rs = 2 G \Mbh / c^2$. As a result, for a given black hole mass, some stellar types may be vulnerable to tidal disruption while others would instead pass through the MBH horizon whole.  Of particular interest to this study is the fact that MBHs more massive than $\sim 10^5 M_\sun$ swallow typical white dwarfs (WDs) whole, while those of lower mass can produce tidal disruptions of WDs. 

Tidal disruptions of WDs, therefore, uniquely probe the long-debated existence of MBHs  with masses less than $10^5 M_\odot$. 
The kinematic traces of such black holes are difficult to resolve spatially due to their relatively small radii of gravitational influence, even with the {\it Hubble Space Telescope}, which has proven a powerful tool for probing more massive nuclei \citep[e.g.][]{Lauer:1995dm,Seth:2010kh}. 
While current observational constraints suggest that black holes are ubiquitous in giant galaxies  \citep{Richstone:1998wk}, their presence is more uncertain in dwarf galaxies \citep[although, see][]{Reines:2011cs}. 
Determination of the galactic center black hole mass function has traditionally focused on active galaxies  \citep[e.g.][]{Kelly:2012dj,Miller:2014vi}, for which we can directly infer the black hole mass,
and work by \citet{Greene:2004cz}, \citet{Greene:2007bq,Greene:2007dz} and \citet{Reines:2013bp} has shown intriguing possibilities for active galactic center black holes with masses  similar to $10^5 M_\odot$. 
But, observations of tidal disruption events probe the mass function of otherwise quiescent black holes, offering a powerful check on mass functions derived from their active counterparts \citep{Gezari:2009dna}. 

With this motivation, predicting the signatures of tidal interactions between WDs and MBHs has been the subject of substantial  effort. Studies find that the resulting mass transfer nearly always exceeds the MBH's Eddington limit mass accretion rate, $\dot M_{\rm Edd} = 2 \times 10^{-3} (\Mbh/10^5M_\odot)\ M_\odot \text{ yr}^{-1}$, where we have used $L_{\rm Edd} = 4\pi G \Mbh m_{\rm p} c / \sigma_{\rm T}$ and a 10\% radiative efficiency, $L=0.1 \dot M c^2$.
Accretion disk emission from these systems is at most $\sim L_{\rm Edd}$, which is increasingly faint for smaller MBHs \citep[e.g.][]{Beloborodov:1999wb}. 
However, we expect that these systems also launch relativistic jets as a result of the extremely rapid mass supply \citep{2011MNRAS.416.2102G,Krolik:2012da,DeColle:2012bq,Shcherbakov:2013hf}.
The observed luminosity of these jetted transients is likely to be proportional to $\dot M c^2$ \citep{DeColle:2012bq} and thus may greatly exceed $L_{\rm Edd}$ when $\dot M \gg 
\dot M_{\rm Edd}$. While disk emission may peak at ultraviolet or soft x-ray frequencies \citep{RamirezRuiz:2009gw,Rosswog:2009gg}, 
the jetted emission can be either produced by internal dissipation or by  Compton-upscattering  the disk photon field  to higher frequencies \citep[e.g.][]{Bloom:2011er,Burrows:2011kz}. We turn our attention to these luminous high-energy jetted transients arising from WD-MBH interactions in this paper. 

Despite the general feature of high accretion rates, theoretical studies predict a wide diversity of signatures depending on the orbital parameters with which the WD encounters the MBH. 
Single, strongly-disruptive passages are thought to produce quick-peaking lightcurves with power-law decay tails as debris slowly falls back to the MBH \citep{Rosswog:2009gg,Haas:2012ci,Cheng:2013cm,Shcherbakov:2013hf}. 
It has also been suggested that these sufficiently deeply-passing encounters may result in detonations of the WD \citep{Luminet:1989wl,Rosswog:2009gg,Haas:2012ci,Shcherbakov:2013hf,Holcomb:2013et},
and thereby accompany the accretion flare with a simultaneous type I supernova \citep{Rosswog:2008gc}.   
Multiple passage encounters result in lightcurves modulated by the orbital period \citep{Zalamea:2010eu,2013ApJ...777..133M}, and
recent work by \citet{AmaroSeoane:2012fl}, \citet{Hayasaki:2013kd}, and \citet{Dai:2013un} has shown that the mass fallback properties from eccentric orbits should be quite different from those in near-parabolic encounters. 
 \citet{Krolik:2011ew} have suggested that tidal stripping of a WD might explain the variability in the lightcurve of {\it Swift} J1644+57 \citep{Levan:2011hq,Bloom:2011er,Burrows:2011kz}. 
Finally, \citet{Dai:2013dn} and \citet{Dai:2013bt} have shown that the Roche lobe overflow of a WD in a circular orbit around a MBH will produce stable mass transfer, and a long-lived accretion flare.  Transients in which the WD completes many orbits are of particular interest as they are persistent gravitational radiation sources with  simultaneous electromagnetic counterparts \citep{Sesana:2008ii}.

We review the properties of transients produced by tidal interactions between WDs and MBHs, with particular emphasis on the role that the orbit may play in shaping the ensuing mass transfer from the WD to the MBH in Section \ref{sec:sig}. 
We focus on cases where the supply of material to MBH is above the hole's Eddington limit and launches a relativistically-beamed jet component.  
In Section \ref{sec:rates}, we discuss our assumptions about the nature of stellar clusters surrounding MBHs. 
We model the tidal and gravitational wave-driven capture of WDs into bound orbits in order to predict the orbital distribution and rates of eccentric and circular mass transfer scenarios in Section \ref{sec:inspiral}. 
We find that these events are likely outnumbered by single-passage disruptions.
In Section \ref{sec:detection}, we illustrate  that although they are rare, WD disruptions may sufficiently outshine MS disruptions in jetted transients that they should be easily detectible. In Section \ref{sec:discussion}, we argue that the detection or non-detection of these transients should place strong limits on the existence of MBHs with masses less than $10^5 M_\odot$. Finally, 
we show that WD-MBH interaction transients bear similarities in peak luminosity and timescale to the newly-identified ultra-long gamma ray bursts  \citep[GRBs;][]{Levan:2014iz}.

\section{Phenomenology of White Dwarf Tidal Interactions}\label{sec:sig}

We can distinguish between WD-MBH encounters based on the orbit with which the WD begins to transfer mass to the MBH. This section reviews some of the expected signatures of these encounters, emphasizing the role of the orbital eccentricity at the onset of mass transfer. 

\begin{figure}[tbp]
\begin{center}
\includegraphics[width=0.44\textwidth]{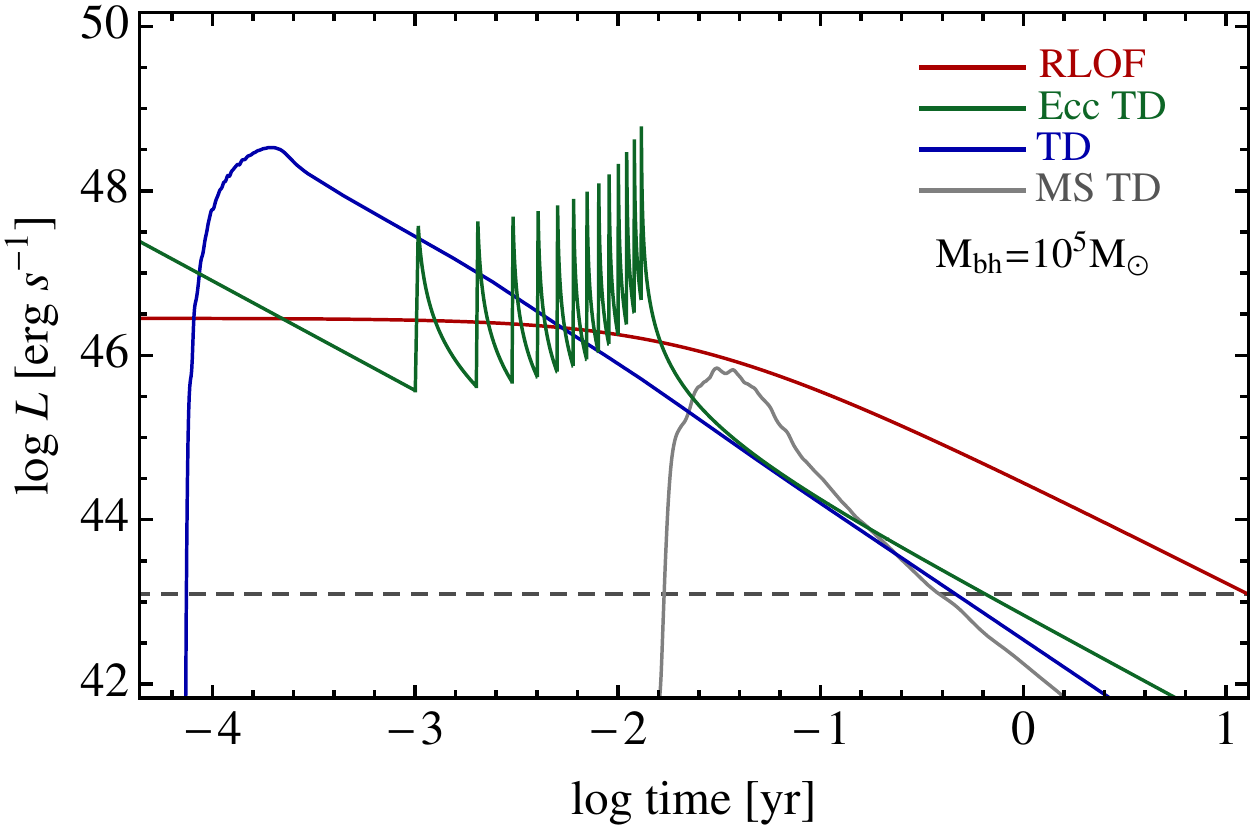}
\caption{ Accretion-powered flares that result from tidal interactions between $0.5 M_\odot$ WDs and a $10^5 M_\odot$ MBH, calculated assuming that $L = 0.1 \dot M c^2$.  A tidal disruption event with $r_p = r_t$ is shown in blue, a repeating flare due to tidal stripping of the WD in an eccentric orbit is shown in green, and Roche lobe overflow (RLOF) and the ensuing stable mass transfer is shown in red.  For comparison, the gray line shows disruption of a sun-like star and the dashed line shows the Eddington luminosity for a $10^5 M_\odot$ black hole. Tidal disruption $\dot M(t)$ curves are from \citet[][available online at astrocrash.net]{Guillochon:2013jj}. A wide diversity of flare characteristics are achieved with differing orbital parameters.     }
\label{fig1}
\end{center}
\end{figure}

In Figure \ref{fig1}, we show representative light curves,  calculated by assuming that $L=0.1 \dot M c^2$ for each of the orbital classes we will consider below. Transients produced range from a slow, smooth decline for Roche lobe overflow to multiple short-timescale flares in the eccentric tidal stripping case. 
We presume a WD mass of $0.5 M_\odot$ and a MBH mass of $10^5 M_\odot$ in Figure \ref{fig1} \citep[see e.g.][for discussions of the single and binary WD mass distributions]{Kepler:2007jz,Maoz:2012dp}.  In all of the following we will assume that the WD mass radius relationship is described by 
\beq\label{eq:massradius}
R_{\rm wd} = 0.013 R_\odot \left( \frac{1.43 M_\odot}{M_{\rm wd}} \right)^{1/3} \left( 1- \frac{M_{\rm wd}}{1.43 M_\odot} \right)^{0.447},
\eeq
from \citet{Zalamea:2010eu}. Where relevant, we will further assume that the internal structure of the WDs is described by that of a $n=3/2$ polytrope \citep[e.g.][]{Paschalidis:2009hk}. This is strictly most relevant at low WD masses, but because low-mass white dwarfs are the most common \citep{Maoz:2012dp} and also those most vulnerable to tidal interactions, we suggest this may be a reasonable approximation for most astrophysically relevant cases.

\subsection{Near-parabolic orbit tidal disruption}
Typical tidal disruption events occur when stars are scattered by two-body relaxation processes in orbital angular momentum into orbits that pass close to the black hole at pericenter. We will parameterize the strength of the encounter with $\beta\equiv\rt/\rp$, such that higher $\beta$ correspond to deeper encounters as compared to the tidal radius.  Simulations of WD disruptions  have been performed recently by several authors \citep{Rosswog:2008gc,Rosswog:2009gg,Haas:2012ci,Cheng:2013cm}, and we describe some of the salient features here. 

The vast majority of these orbits originate from quite far from the MBH, where star-star scatterings become substantial \citep{Frank:1978wx,Merritt:2013ta}. As a result, typical orbits are characterized by $e\approx1$.  The aftermath of a disruption has been well determined in the limit that the spread in binding energy across the star at pericenter is large compared to its original orbital energy \citep{Rees1988}. The critical orbital eccentricity above which the parabolic approximation holds is
\citep{AmaroSeoane:2012fl,Hayasaki:2013kd}
 \beq
 \label{ecrit}
e > e_{\rm crit} \approx 1-{2\over \beta} \left( {M_* \over \Mbh }\right)^{1/3}.
\eeq  For a $\beta =1$ encounter between a $0.5 M_\odot $  WD and a $10^5 M_\odot$ MBH, $e_{\rm crit} \approx 0.97$. 

If $e > e_{\rm crit}$, about half of the debris of tidal disruption is bound to the MBH, while the other half is ejected on unbound orbits \citep{Rees1988,Rosswog:2008hv}. The initial fallback of the most bound debris sets the approximate timescale of peak of the lightcurve, which scales as $\tfb \propto \Mbh^{1/2}M_*^{-1}R_*^{3/2}$. The peak accretion rate, which is proportional to $\dot M_{\rm peak} \propto \Delta M / \tfb$, thus scales as  $\dot M_{\rm peak} \propto \Mbh^{-1/2}M_*^{2}R_*^{-3/2}$ \citep{Rees1988}.  The fallback curves typically feature a fast rise to peak, and then a long, power-law decay with asymptotic slope similar to $t^{-5/3}$ \citep[though, see][]{Guillochon:2013jj}. Since the orbital time at the tidal radius is much shorter than that of the most bound debris, it is usually assumed that the accretion rate onto the MBH tracks the rate of fallback \citep{Rees1988}. 

\begin{figure*}[tbp]
\begin{center}
\includegraphics[width=0.95\textwidth]{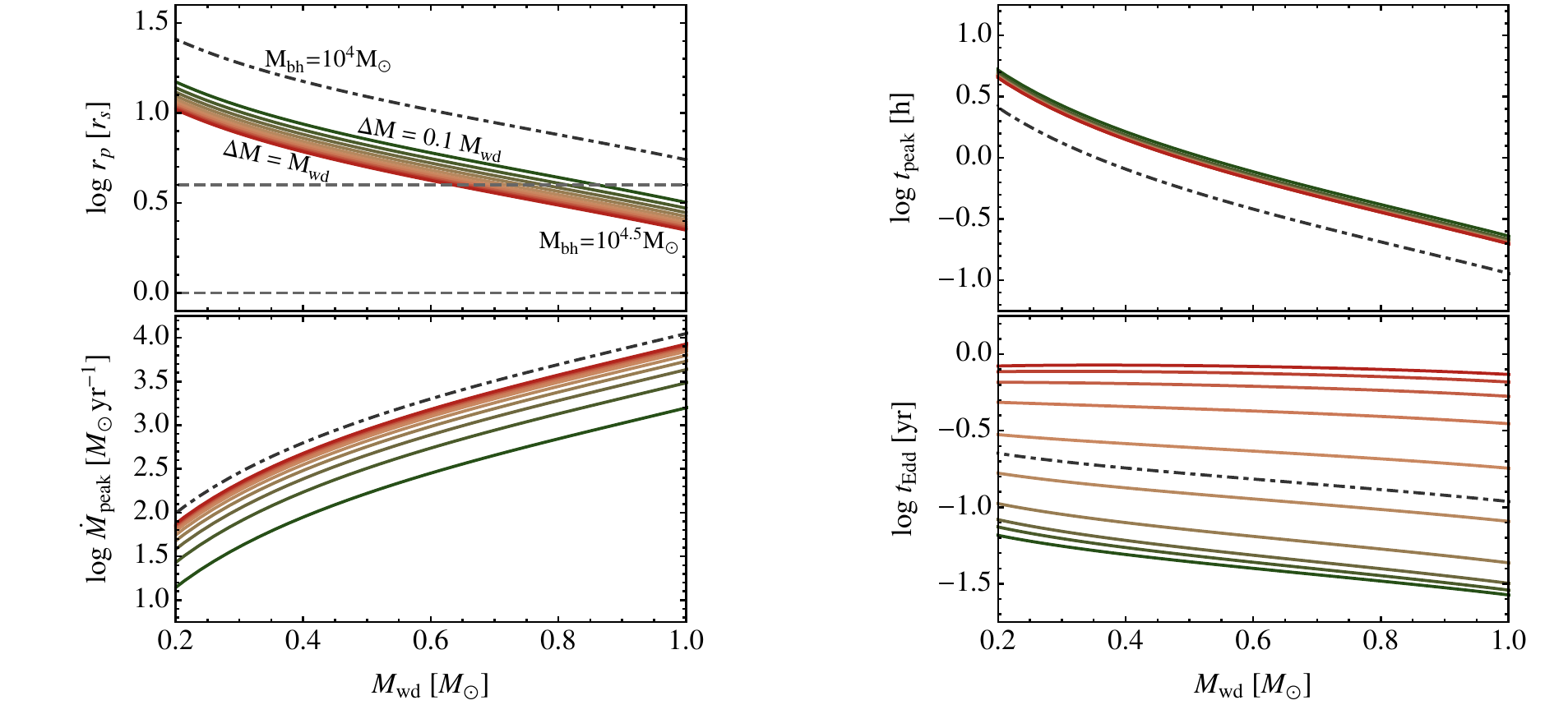}
\caption{ The properties of tidal disruptions of WDs with masses $0.2-1.0 M_\odot$ encountering MBHs with masses of $10^4$ and $10^{4.5} M_\odot.$ Colored lines represent encounters with a $10^{4.5} M_\odot$ MBH in which the WD loses a fraction between 0.1 and 1 of its total mass, in intervals of 0.1. Dot-dashed lines represent encounters in which half of the WD mass is stripped in an encounter with a $10^4 M_\odot$ MBH. 
The upper left panel shows that disruptive encounters occur outside the MBH's Schwarzschild radius for the range of masses considered, but many close passages have $\rp<r_{\rm ISCO}$, which may be a more appropriate cutoff for determining whether an accretion flare or prompt swallowing results from a given encounter. The remaining panels draw on simulation results from \citet{Guillochon:2013jj} for $n=3/2$ polytropes to show the peak $\dot M$, timescale of peak, $t_{\rm peak}$, and time spent above the Eddington limit, $t_{\rm Edd}$.  } 
\label{fig2}
\end{center}
\end{figure*}

In Figure \ref{fig2}, we estimate typical properties for encounters between WDs of various masses and MBHs of $10^4$ and $10^{4.5} M_\odot$. To construct this Figure, we draw on results of hydrodynamic simulations of tidal disruption of $n = 3/2$ polytropic stars performed by \citet{Guillochon:2013jj}. We plot colored lines corresponding to ten different impact parameters, where the WD would lose a fraction $0.1 - 1$ of its mass in intervals of 0.1 in an encounter with a $10^{4.5} M_\odot$ MBH. We plot a single dot-dashed line for a 50\% disruptive encounter between a WD and a MBH.  All of these events fuel rapid accretion to the MBH with typical accretion rates ranging from hundreds to thousands of solar masses per year. Typical peak timescales for the accretion flares are hours. The long-term fallback fuels accretion above the Eddington limit for a period of months, after which one might expect the jet to shut off, terminating the high-energy transient emission \citep{DeColle:2012bq}.

In the upper left panel, we compare pericenter distance to both $\rs$ and $r_{\rm ISCO} \approx 4\rs$. Simulations of tidal encounters in general relativistic gravity, for example those of \citet{Haas:2012ci} and \citet{Cheng:2013cm} indicate that if the pericenter distances $\rp \sim \rs$, relativistic precession becomes extremely important and free-particle trajectories deviate substantially from Newtonian trajectories. We expect, therefore, that encounters with  $\rp \lesssim r_{\rm ISCO}$ will experience strong general relativistic corrections to the orbital motion of tidal debris. The result is likely to be prompt swallowing of the bulk of the tidal debris rather than circularization and a prolonged accretion flare. For that reason we will use $r_{\rm ISCO}$ as a point of comparison for determining when stars are captured whole or produce a tidal disruption flare in this paper \citep[as suggested, for example in chapter 6 of ][]{Merritt:2013ta}. Future simulations of these extreme encounters will help distinguish where the exact cutoff between capture and flaring lies.

\subsection{Tidal stripping in an eccentric orbit}
From an eccentric orbit, if $e<e_{\rm crit}$, equation \eqref{ecrit}, all of the debris of tidal disruption is bound to the MBH \citep{AmaroSeoane:2012fl,Hayasaki:2013kd,Dai:2013un}. If it is only partially disrupted, the remnant itself will return for further passages around the MBH and perhaps additional mass-loss episodes \citep{Zalamea:2010eu}. 
We explore the nature of the accretion that results from the progressive tidal stripping of a WD in this section. In Sections \ref{sec:rates} and \ref{sec:inspiral}, we will elaborate on the stellar dynamical processes that can lead a WD to be captured into such an orbit. 

Tightly-bound orbits around the MBH are very well described by Keplerian motion in the MBH's gravitational field and the WD should be only scattered weakly each orbit. In other words, its orbital parameters diffuse slowly in response to any background perturbations (see Section \ref{sec:rates} for more discussion of the stellar dynamics of tightly-bound stars) \citep[e.g.][]{Hopman:2005ij}. Thus, when such a WD first enters the tidal radius, it would do so only grazingly, losing a small fraction of its mass. We suggest in Sections \ref{sec:rates} and \ref{sec:inspiral} that a more typical process may be progressive tidal forcing to the point of disruption. In this picture, a WD in an initially non-disruptive orbit is eventually disrupted by the build up of tidal oscillation energy. Over many passages orbital energy is deposited into $l=2$ mode oscillation energy of the WD \citep{Baumgardt:2006fg}. Eventually, the oscillation energy exceeds the WD's gravitational binding energy and mass is stripped from the WD envelope.

After the onset of mass transfer between the WD and the MBH, the WD will expand in radius and decrease in density following equation \eqref{eq:massradius}. The strength of subsequent encounters increases until the WD is completely destroyed by the MBH. At each encounter, we calculated the new $\beta$ parameter based on the adjusted mass, and in turn the corresponding $\Delta M$. The exact extent of mass loss may be modulated through the superposition of the WD's oscillation phase and tidal forcing at pericenter \citep{Mardling:1995hx,Mardling:1995it,Guillochon2011}.  Unlike, for example, a giant star being tidally stripped \citep[e.g.][]{2013ApJ...777..133M}, as long as degeneracy is not lifted the internal structure of the WD remains polytropic. We find that, over the course of tens of orbits, the mass loss episodes escalate from $<10^{-2}M_\odot$ until the remaining portion of the WD is destroyed. This is in contrast to the calculation of \citet{Zalamea:2010eu}, who, as a result of using a more approximate formula for $\Delta M (\beta)$ with shallower $\beta$-dependence, predict that the tidal stripping episode will persist for $\sim 10^4$ orbits. If additional heating occurs near the surface of the WD due to interaction between oscillations and marginally-bound material, as, for example, observed in simulations of WD  \citep{Cheng:2013cm} and giant-star disruptions \citep{2013ApJ...777..133M}, the degeneracy of the outermost layers of the WD may be lifted, leading to an even more rapid exponentiation of mass-loss episodes. 

The example in Figure \ref{fig1} shows a WD being stripped in an orbit with a period of $10^{4.5}$ seconds. This timescale sets the repetition time of the flares, and corresponds to $e\approx0.97\approx e_{\rm crit}$. One consequence of orbits with lower eccentricity is that the fallback of the bound material happens on very rapid timescales, potentially more rapidly than material that circularizes at the tidal radius may be viscously accreted. The ratio of fallback time (here estimated by the orbital period) to viscous time at pericenter is approximately,
\beq
\frac{t_{\rm visc}}{t_{\rm orb}} \approx 2 \left( \frac{1-e}{0.03} \right)^{3/2} \left( \frac{\alpha_{\nu}}{0.01} \right)^{-1} \left( \frac{H/R}{0.5} \right)^{-2},
\eeq 
where $\alpha_{\nu}$ is the Shakura-Sunayev viscosity parameter \citep{Shakura:1973uy}.

When the viscous time is longer than the fallback time, the lightcurve will represent the viscous accretion of the nearly-impulsively assembled torus of tidal debris. 
To illustrate the accretion rate that may be expected, we employ a super-Eddington disk model proposed by \citet{Cannizzo:2009gm} and employed by \citet{Cannizzo:2011df} to describe super-Eddington accretion in the case of {\it Swift} 1644+57. In this simple model,  the accretion rate of an impulsively assembled disk is mediated by the rate of viscous expansion of the material. Quoting from \citet{Cannizzo:2011df}, the peak accretion rate is 
\beq
\begin{aligned}
\dot M_0 &=& 273 \left(  \frac{\Delta M}{10^{-2} M_\odot } \right)  \left(  \frac{\Mbh}{10^5 M_\odot } \right)^{1/2} \times \\
&&  \left(  \frac{\rp}{10^{11}\text{cm} } \right)^{-3/2}   \left(  \frac{\alpha_\nu}{0.01 } \right) M_\odot \text{ yr}^{-1}.
\end{aligned}
\eeq
The behavior in time is then
\beq
\dot M(t) = \dot M_0 \left( \frac{t}{t_0} \right)^{-4/3},
\eeq
where $t_0 = 4/9 \alpha_\nu^{-1} \rp^{3/2}(G \Mbh)^{-1/2}$, approximately the viscous time at pericenter for a thick disk. The $t^{-4/3}$ proportionality is similar to that derived in the case of zero wind losses for impulsively assembled disks by \citet{Shen:2013vo}. \citet{Shen:2013vo} go on to show if the fraction of material carried away in a wind is, for example, 1/2, then the time proportionality steepens to $-5/3$. This indicates that the $t^{-4/3}$ power-law decay plotted in Figure \ref{fig1} is at the shallow end of the range of possible behavoir. Any degree of wind-induced mass loss from the disk would steepen the slope of the falloff in time, further reducing the luminosity between flaring peaks.  \citet{Coughlin:2014fs} have recently proposed a new thick disk and jet launching model for super-Eddington accretion phases of tidal disruption events. Their ZEBRA model can capture the rise and peak phases in the lightcurve, not just the late-time decay behavior. These characteristics will be essential in making constraining comparisons to potential future observations.

\subsection{Roche-lobe overflow from a circular orbit }
If the WD reaches the tidal radius in a circular orbit, mass transfer will proceed stably \citep{Dai:2013bt}. The rate at which gravitational radiation carries away orbital angular momentum,
\beq
\dot J_{\rm GR} = - \frac{32}{5} \frac{G^3}{c^5} \frac{\Mbh M_{\rm wd} \left( \Mbh + M_{\rm wd}  \right) }{a^4} J_{\rm orb}
\eeq
is balanced by the exchange of mass from the WD to MBH, and the corresponding widening of the orbit \citep[e.g.][]{Marsh:2004fv}. The resulting equilibrium mass transfer rate is then given by 
\beq
\dot M_{\rm wd}  = \left[1+ \frac{\zeta_{\rm wd} - \zeta_{\rm rl}}{2}  - \frac{M_{\rm wd}}{\Mbh} \right]^{-1} \frac{\dot J_{\rm GR} } {J_{\rm orb}} M_{\rm wd}
\eeq
where $\zeta_{\rm wd}$ and $\zeta_{\rm rl}$ are the coefficients of expansion of the WD and Roche lobe, respectively, in response to mass change, $\zeta = d\ln r/d \ln M$, \citep[eq. 19 in][]{Marsh:2004fv}. For low mass WDs, $\zeta_{\rm wd} \approx -1/3$, while the Roche lobe is well described by $\zeta_{\rm rl} \approx 1/3$.

As a result of the stability, mass transfer between a WD and a MBH would persist above the Eddington limit for multiple years. They would also radiate a persistent gravitational wave signal, with frequencies of order $\sim (G \Mbh / \rt^3)^{1/2}$, or about $0.2$Hz for a $10^5 M_\odot$ MBH and $0.5 M_\odot$ WD. Such frequencies would place these objects within the sensitivity range of the proposed {\it LISA} and {\it eLISA} missions \citep[e.g.][]{Hils:1995en,Freitag:2003kg,Barack:2004jy,AmaroSeoane:2007im}.

\section{Stellar Clusters Surrounding MBHs}\label{sec:rates}

The properties of the stellar systems that surround MBHs determine the the orbital parameters with which WDs encounter MBHs. The nature of the stellar systems that surround MBHs with masses less than $10^6 M_\odot$ remains observationally unconstrained. However, dense stellar clusters appear to  almost universally surround known galactic center MBHs, which typically span the mass range of $10^6$-$10^9M_\odot$.  Even in galaxies that lack nuclear activity, dense stellar clusters in galactic nuclei with centrally-peaked velocity dispersion profiles  strongly suggest the presence of central massive objects \citep[e.g.][]{Lauer:1995dm,Byun:1996hw,Faber:1997fn,Magorrian:1998cs}.
That MBHs should be surrounded by stars is not entirely unexpected. With a mass much greater than the average mass of surrounding stars, a MBH sinks rapidly to the dynamical center of the stellar system in which it resides \citep{Alexander:2005ij}.  There may also exist a population of nearly ``naked"  MBHs only surrounded by a  hyper-compact stellar cluster \citep{Merritt:2009es,OLeary:2009ga,OLeary:2012hy,Rashkov:2013hl,Wang:2014vp}. Such systems originate in dynamical interactions that lead to the high velocity ejection of MBHs from their host nuclei.

\subsection{A Simple Cluster Model}\label{sec:clmodel}
\begin{figure}[tbp]
\begin{center}
\includegraphics[width=0.45\textwidth]{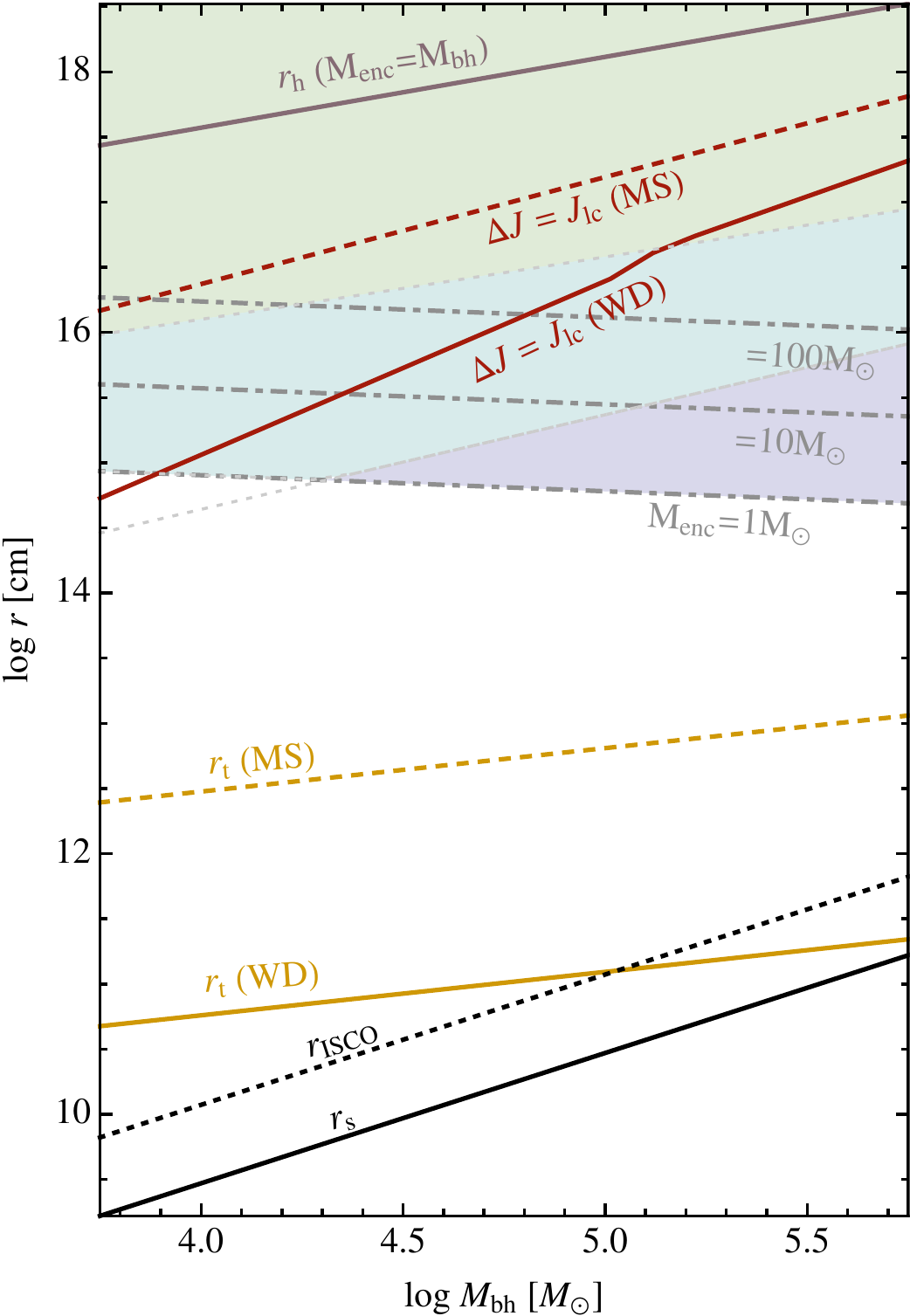}
\caption{ Characteristic scales for WD-MBH interactions in stellar cusps surrounding MBHs given the cluster properties described in Section \ref{sec:clmodel}. Shown, from bottom to top, are: 1) the Schwarzschild radius, $\rs$, and the radius of the Innermost Stable Circular Orbit $\sim 4 \rs$ (black solid and dotted, respectively), 2) the tidal radius, $\rt$ for WDs $(0.5M_\odot)$ and MS (sun-like) stars (yellow solid and dashed), 3) the radii that enclose 1, 10, and 100 $M_\odot$ (gray dot-dashed), 4) The characteristic orbital semi-major axis that marks the transition from the empty (smaller $a$) to the full (larger $a$) loss cone regimes (red solid and dashed lines labeled $\Delta J = \Jlc$), and 5) the MBH radius of influence, $r_{\rm h}$.   Filled shading denotes the region that is, on average, populated by stars. Filling colors denote the primary orbital relaxation mechanism with general-relativistic resonant relaxation (purple), mass-precession resonant relaxation (cyan), and finally non-resonant relaxation (green) being dominant from small to large radii, respectively.     }
\label{fig:clscales}
\end{center}
\end{figure}

In what follows, we adopt a simplified stellar cluster model in which the gravitational potential is Keplerian (dominated by the black hole), and the stellar density is a simple power-law with radius. Our approach is very similar to that of \citet{MacLeod:2012cd}. 
In Figure \ref{fig:clscales}, and in the following paragraphs, we introduce the relevant scales that describe the orbital dynamics of such a system.

MBHs embedded in stellar systems are the dominant gravitational influence over a mass of stars similar to their own mass. 
  At larger radii within the galactic nucleus, the combined influence of the MBH and all of the stars describes stellar orbits. 
Keplerian motion around the MBH is energetically dominant within the MBH's radius of influence 
 \beq\label{rh}
r_{\rm h} = \frac{G M_{\rm bh}}{\sigma_{\rm h}^2}  = 0.43 \left(\frac{M_{\rm bh}}{10^5 M_\odot }\right)^{0.54} \ {\rm pc},
\eeq
where $\sigma_{\rm h}$ is the velocity dispersion of the surrounding stellar system.  We will assume that the velocity dispersion of stellar systems surrounding MBHs can be approximated by the $\Mbh-\sigma$ relation  \citep[e.g. ][]{Ferrarese2000,Gebhardt2000,Tremaine:2002hd,Gultekin:2009hj,Kormendy:2013vg}. 
This assumption, by necessity, involves extrapolating the $\Mbh-\sigma$ relation to  lower black hole masses than those for which it was derived. We've adopted $\sigma_{\rm h} = 2.3 \times 10^5 (M_{\rm bh}/M_\odot)^{1/4.38}  \ {\rm cm \ s^{-1}}$  \citep{Kormendy:2013vg}.

To normalize the mean density of the stellar cluster, we assume that the enclosed stellar mass within $r_{\rm h}$ is equal to the MBH mass, $ M_{\rm enc}(r_{\rm h})  = \Mbh $.
Despite the uncertainty in extrapolating the $\Mbh-\sigma$ relation \citep[e.g.][]{Graham:2013fc}, this exercise can provide a telling estimate of the order of magnitude rates of interactions between WDs and MBHs should the $\Mbh-\sigma$ relation actually extend to lower masses. This calculation more robustly constrains the WD interaction rate relative to other interactions that are also based on the density of the stellar cluster, like main-sequence star disruptions.

In energetic equilibrium, stars within this radius of influence distribute according to a power-law density profile in radius.  We will show following equation \eqref{tNRR} that the energetic relaxation time for stellar clusters (of the masses we consider) is short compared to their age. Thus, the assumption of an equilibrated stellar density profile is realistic. The slope of this power-law depends on the mass of stars considered as compared to the average stellar mass \citep{Bahcall:1976kk,Bahcall:1977ea}. We adopt a stellar number density profile $\nu_* \propto r^{-\alpha}$ with $\alpha=3/2$ in Figure \ref{fig:clscales} and the examples that follow.  As a result, the enclosed mass as a function of radius is
$ M_{\rm enc}  = \Mbh \left(r/r_{\rm h}\right)^{3/2} $. If we assume that the angular momentum distribution is isotropic, then this radial density profile also defines the distribution function of stars in orbital binding energy, $\e$, \citep{Magorrian:1999fd},
\beq
f(\e) = \left( 2 \pi \sigma_{\rm h}^2 \right)^{-3/2} \nu_*(r_{\rm h}) \frac{\Gamma \left(\alpha+1\right)}{\Gamma \left(\alpha-{1\over 2} \right)} \left( \e \over \sigma_{\rm h}^2 \right)^{\alpha-{3 \over 2}}.
\eeq
This density profile also sets the local one-dimensional velocity dispersion, 
\beq
\sigma^2 = \frac{G \Mbh}{(1+\alpha) r},
\eeq
in the region $r\ll r_{\rm h}$.

If the outermost radius of the cluster is defined by the radius of influence, then the characteristic inner radius is the distance from the MBH at which the enclosed stellar mass is similar to the mass of a single star. This scale provides insight into the expected binding energy of the most bound star in the system.
As a result, the radius that encloses  a single stellar mass is
\beq\label{rmbar}
r_{\bar m_*} = 
 \left( {  \bar m_* \over \Mbh  } \right)^{2/3} r_{\rm h},
\eeq
where ${\bar m}_*$ is the average stellar mass. 
For simplicity, we adopt $\bar m_* = 1 M_\odot$.  In reality, mass segregation will create a gradient in which the average mass may vary substantially as a function of radius. It is possible that objects as large as $10M_\odot$, for example stellar-mass black holes, may be the dominant component at very small radii. However, because we adopt a radius-independent value of $\bar m_*$, we take $1M_\odot$ as representative of the turnoff mass of a $\sim 12$ Gyr-old stellar population \citep[see][for a more thorough discussion]{Alexander:2005ij}.  We plot the radii that enclose 1, 10, and 100 $M_\odot$ in Figure \ref{fig:clscales}.

\subsection{Orbital Relaxation}
Within a dense stellar system, stars orbit under the combined influence of the MBH and all of the other stars. As a result, their orbital trajectories are constantly subject to perturbations, and deviate from closed, Keplerian ellipses. The magnitude of these perturbations may be estimated by comparing the orbital period, $P$, to the orbital relaxation time, $t_{\rm r}$. 
For most stars, two-body relaxation drives orbital perturbations
\beq\label{tNRR}
t_{\rm NRR} = \frac{0.34 \sigma^3}{G^2 \bar m_* \rho \ln \Lambda},
\eeq
equation 3.2 of \citet{Merritt:2013ta}, also see \citep{Binney2008,Alexander:2005ij}. We adopt a value for the Coulomb logarithm of $\ln \Lambda = \ln \left(\Mbh/\bar m_* \right)$, the natural log of the number of stars within the sphere of influence. Under these assumptions, a cluster with a   $\Mbh = 10^5 M_\odot$ MBH and $\bar m_* = 1M_\odot$ would have undergone approximately 160 relaxation times within the age of the universe.  Only when $\Mbh\approx 10^{6.75}$ does the relaxation time equal the Hubble time, suggesting that the choice of a relaxed, power-law distribution of stellar number density  is appropriate for the MBH masses we consider.

 Tightly bound stellar orbits are also perturbed by secular torques from the orbits of nearby stars. An important aspect of estimating the ``resonant relaxation" evolution time of a star's orbit in response to these torques is estimating the coherence time of the background orbital distribution. The coherence time is the typical timescale on which neighboring orbits precess, and thus depends on the mechanism driving the precession.
When this coherence time is determined by Newtonian advance of the argument of periastron, or mass precession, the incoherent resonant relaxation time is a factor of $ \Mbh / \bar m_* $ greater than the orbital period, 
\beq\label{tRRN}
t_{{\rm RR},M} =  \left(  \Mbh \over \bar m_* \right) P,
\eeq
and $t_{{\rm coh},M} = \Mbh P/M_{\rm enc}$  \citep[equations 5.240 and 5.202 of][]{Merritt:2013ta}. 
The orbital period $P$ is defined as $P(a)$, the Keplerian orbital period for a semi-major axis, $a$, equal to $r$. Where general-relativistic precession determines the coherence time, the incoherent resonant relaxation time is
\beq\label{tRRGR}
t_{{\rm RR},GR} = {3 \over \pi^2} {r_{\rm g} \over a} \left(  \Mbh \over \bar m_* \right)^2 {P \over N_{\rm enc}},
\eeq
for a coherence time of $t_{{\rm coh},GR} =  a P/(12 r_{\rm g})$, where $r_{\rm g} = G \Mbh/c^2$ and $N_{\rm enc} = M_{\rm enc}/ \bar m_*$ \citep[equations 5.241 and 5.204 of][]{Merritt:2013ta}. Either equation \eqref{tRRN} or \eqref{tRRGR} determines the resonant relaxation timescale, $t_{\rm RR}$ depending on which coherence time is shorter.

 We take the relaxation time, $t_{\rm r}$, to be the minimum of the resonant and non-resonant relaxation timescales, 
 \beq
 t_{\rm r} = \min(t_{\rm NRR},t_{\rm RR}). 
 \eeq
The background shading in Figure \ref{fig:clscales} shows the dominant relaxation mechanism as a function of semi-major axis. First general relativistic resonant relaxation \eqref{tRRGR}, then mass-precession resonant relaxation \eqref{tRRN}, and finally non-resonant relaxation \eqref{tNRR} dominate from small to large radii.

\subsection{Scattering to the Loss Cone}
Within a relaxation time, stellar orbits exhibit a random walk in orbital energy and angular momentum. Orbits deviate by of order their energy and the corresponding circular angular momentum in this time \citep[e.g.][]{Lightman:1977hu,Cohn:1978cc,Merritt:2013ta}. Thus, the root-mean-square change in angular momentum per orbit is $\Delta J =   \Jc \sqrt{ P / \tr } $. The characteristic angular momentum of orbits that encounter the black hole is the loss cone angular momentum, 
 $\Jlc \approx \sqrt{ 2 G \Mbh r_{\rm p}  }$, where $r_{\rm p}$ is the the larger of the tidal radius, $\rt$ and the black hole Schwartzschild radius $\rs$ \citep{Lightman:1977hu}.
This loss cone angular momentum is significantly smaller than the circular angular momentum, $\Jlc \ll \Jc$. Thus, the timescale for the orbital angular momentum to change of order the loss cone angular momentum is typically much less than the relaxation time. As a result, stars tend to be scattered into disruptive orbits via their random walk in angular momentum rather than in energy \citep{Frank:1978wx}. 

A comparison between the loss cone angular momentum, $\Jlc$, and the mean scatter, $\Delta J$, gives insight into the ability of orbital relaxation to repopulate the phase space of stars destroyed through interactions with the black hole. Where $\Delta J \gg \Jlc$ the loss cone is often described as full \citep{Lightman:1977hu}. Orbital relaxation easily repopulates the orbits of stars that encounter the black hole. Conversely, where $\Delta J \ll \Jlc$, the loss cone is, on average, empty \citep{Lightman:1977hu}. The transition between the full and empty loss cone regimes is typical of the semi-major axes from which most stars will be scattered to the black hole \citep[e.g.][]{Syer:1999gp,Magorrian:1999fd,Merritt:2013ta}. From Figure \ref{fig:clscales}, we can see that this radius of transition lies well within the MBH radius of influence, where the enclosed mass is small.

The flux of objects into the loss cone, and thus their disruption rate, is calculated based these criteria of whether the loss cone is full or empty at a given radius (or equivelently, orbital binding energy, $\e$). The number of stars in a full loss cone is $N_{\rm lc}(\e)=4\pi^2 f(\e) P(\e) \Jlc^2(\e) d\e$ \citep{Magorrian:1999fd}. The rate at which they enter the loss cone is mediated by their orbital period defines a loss cone flux $\Flc(\e) = F_{\rm full}(\e) = N_{\rm lc}(\e) / P(\e)$. In regions where the loss cone is not full, somewhat fewer objects populate the loss cone phase space and $\Flc(\e) < F_{\rm full}(\e)$ \citep{Cohn:1978cc,Magorrian:1999fd,Merritt:2013ta}. The exact expressions we use for $\Flc$ in the empty loss cone regime are not reprinted here for brevity but are given in equations (24-26) of \citet{MacLeod:2012cd} and come from the model of \citet{Magorrian:1999fd}.
Once the loss cone flux as a function of energy, $\Flc$, is defined, the overall rate may be integrated
\beq\label{lcrate}
\dot N_{\rm lc} = \int_{\e_{\rm min}}^{\e_{\rm max}} \Flc(\e) d\e,
\eeq
where we take the limits of integration to be the orbital binding energy corresponding to $r_{\rm h}$ $(\e_{\rm min})$ and that corresponding to the $M_{\rm enc} = 10 M_\odot$ radius  $(\e_{\rm max})$.

\begin{figure}[tbp]
\begin{center}
\includegraphics[width=0.45\textwidth]{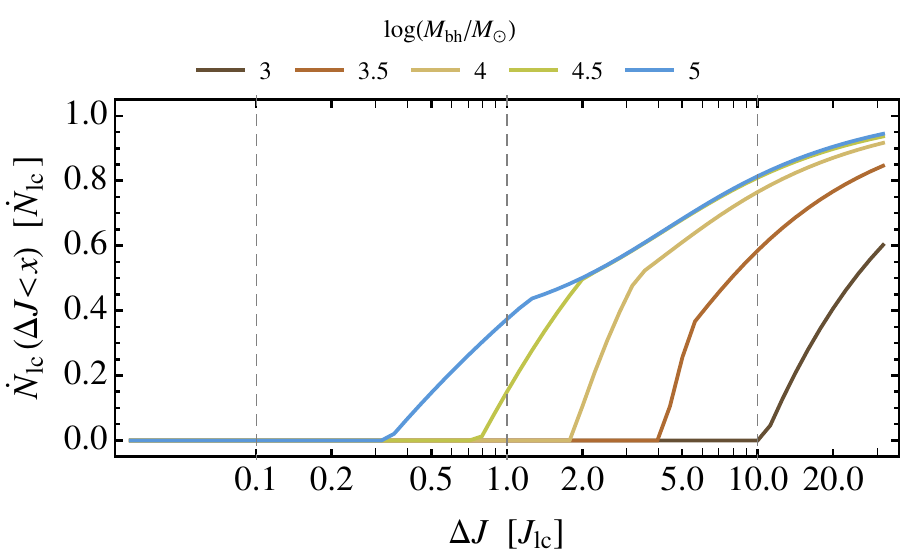}
\caption{ Fraction of the total loss cone flux for which the mean $\Delta J < x$, or the cumulative distribution function of loss cone flux with respect to $\Delta J$. This is computed following equation \eqref{lcrate} where we set $\e_{\rm max}$ based on $\Delta J$ and we assume $0.5M_\odot$ WDs.  We plot lines for MBH masses between $10^3$ and $10^5 M_\odot$. For all black hole masses shown, 100\% of the loss cone flux originates from regions in the cluster where $\Delta J \gtrsim 0.3 \Jlc$, and the bulk of $\dot N_{\rm lc}$ originates from the full loss cone regime where $\Delta J \gg \Jlc$. Because stars receive substantial per-orbit scatter, it is unlikely that they will complete multiple close passages by the MBH with $J\approx\Jlc$.  }
\label{fig:scatter}
\end{center}
\end{figure}

For MBH masses that can disrupt WDs (where  $\rt > r_{\rm ISCO} \approx 4 \rs$), most of the typically populated orbits are in the full loss cone limit. 
Most WDs, therefore, are scattered toward MBHs with mean $\Delta J \gtrsim \Jlc$. In Figure \ref{fig:scatter}, we plot the cumulative distribution function of $\dot N_{\rm lc}$ with respect to $\Delta J/\Jlc$. This shows that for MBHs with $\Mbh < 10^5 M_\odot$, all WDs that reach the loss cone have mean $\Delta J > 0.3 \Jlc$. This has implications for the ability of objects to undergo multiple passages with $J \approx \Jlc$ as we discuss in the next section.

\section{WD Capture and Inspiral}\label{sec:inspiral}

Here, we focus on the stellar dynamics of the capture of WDs into tightly bound orbits from which they can transfer mass to the MBH. We show that WDs are placed into tightly bound orbits primarily through binary splitting by the MBH \citep{2005ApJ...626L..41M}. These orbits then evolve under the influence of tides and gravitational radiation until the WD begins to interact with the MBH.  In modeling this process, we adopt aspects of the pioneering work by \citet{Ivanov:2007fh}. 

\subsection{Binary Splitting and WD Capture}

A key requirement for stars to undergo multiple-passage interactions with a MBH is that the per orbit scatter in angular momentum be sufficiently small that the pericenter distance remains similar between passages. A star in the full loss cone limit $\left(\Delta J \gg \Jlc \right)$ that survives an encounter is very likely to be scattered away from its closely-plunging orbit before it undergoes another encounter.  As we demonstrate in the previous section and in Figure \ref{fig:scatter}, most WDs are in regions where the per-orbit scatter is large relative to $\Jlc$.

Instead, in this section, we focus on the disruption of binary stars scattered toward the MBH, which can leave one star tightly bound to the MBH while the other is ejected on a hyperbolic orbit \citep{Hills:1988br}. Disruptions of binary stars lead WDs to be deposited into orbits from which they are hierarchically isolated from the remainder of the stellar system \citep{AmaroSeoane:2012fl}.  These hierarchically isolated objects have an orbital semi-major axis that is smaller than the region that typically contains stars, $a < r_{\bar m_*}$.  This is the region inside the shaded region of Figure \ref{fig:clscales} as determined by equation \eqref{rmbar}. 
Given such an initial orbit, the WD  may undergo many close passages with the MBH without suffering significant scattering from the other cluster stars.

To estimate the rates and distribution of captured orbits, we follow the analytic formalism of \citet{Bromley:2006fz}, which is motivated by results derived from three-body scattering experiments. \citet{Bromley:2006fz} equations (1)-(5) describe the probability of splitting a binary star as a function of impact parameter, as well as the mean and dispersion in the velocity of the ejected component.  We use these expressions to construct a Monte Carlo distribution of binary disruptions. We let the binaries be scattered toward the black hole with rate according to their tidal radius,  $r_{\rm t, bin} = \left(\Mbh/m_{\rm bin}\right)^{1/3} a$, where $m_{\rm bin}$ is the mass of the binary. We use WD masses of $0.5 M_\odot$ and companion masses of $1M_\odot$ in this example.  We let the binaries originate from the same stellar density distribution described in Section \ref{sec:rates}, with a radially-constant  binary fraction of $f_{\rm bin}=0.1$. For simplicity, we distribute this population of binaries such that there is an equal number in each decade of semi-major axis $dN/da \propto a^{-1}$, within a range $-3 < \log(a/\text{au}) <-1$, although see \citet{Maoz:2012dp} for a more detailed consideration of the separation distribution of field WD binaries. In our simulations, the most tightly bound binaries contribute most to the population that evolves to transfer mass the the MBH, and thus this limit most strongly affects the normalization of our results. The distribution of pericenter distances is chosen given a full loss cone of binaries, such that $dN/d\rp = \text{constant}$,  and we ignore the small fraction of events with $\rp < a$.  

A sampling prior is placed based on the likelihood of a particular encounter occurring. This is estimated by integrating the flux of binaries to the loss cone from the portion of the cluster for which the full loss cone regime applies. 
 This rate is $f_{\rm bin} f_{\rm WD}$ times the nominal loss-cone flux, $\Flc$, integrated from the radius of transition between the full and empty loss cone regimes for a given binary separation outward to $r_{\rm h}$.  This calculation is done following equation \eqref{lcrate} with $\e_{\rm max}$ determined by the binding energy at which $\Delta J = \Jlc$.   Binaries that diffuse toward the black hole gradually from the empty loss cone regime are more likely to undergo a complex series of multiple encounters, the outcome of which is less easily predicted in an analytical formalism \citep[e.g.][]{Antonini:2010cl,Antonini:2011ia}. Therefore we do not include the diffusion of binaries toward the black hole from the empty loss cone regime in our estimate. 

If the captured star has sufficiently small semi-major axis, it will be hierarchically separated from the rest of the steller cluster as the most bound star. The requirement for this condition is that 
\beq\label{aheir}
a < r_{\bar m_*}, 
\eeq
  where $r_{\bar m_*}$ is given by equation \eqref{rmbar} and $\bar m_*=1M_\odot$. When selecting orbits that may undergo many passages, we require them to be hierarchically isolated following equation \eqref{aheir}.   As can be seen in Figure \ref{fig:clscales}, less bound stars (in the full loss cone regime) are subject to major perturbations $\Delta J \gtrsim \Jlc$ each orbit, and thus could not undergo a multiple passage encounter with the MBH. The most tightly bound star, by contrast, evolves in relative isolation from the rest of the cluster until it is subject to a major disturbance. 
 
\begin{figure}[tbp]
\begin{center}
\includegraphics[width=0.42\textwidth]{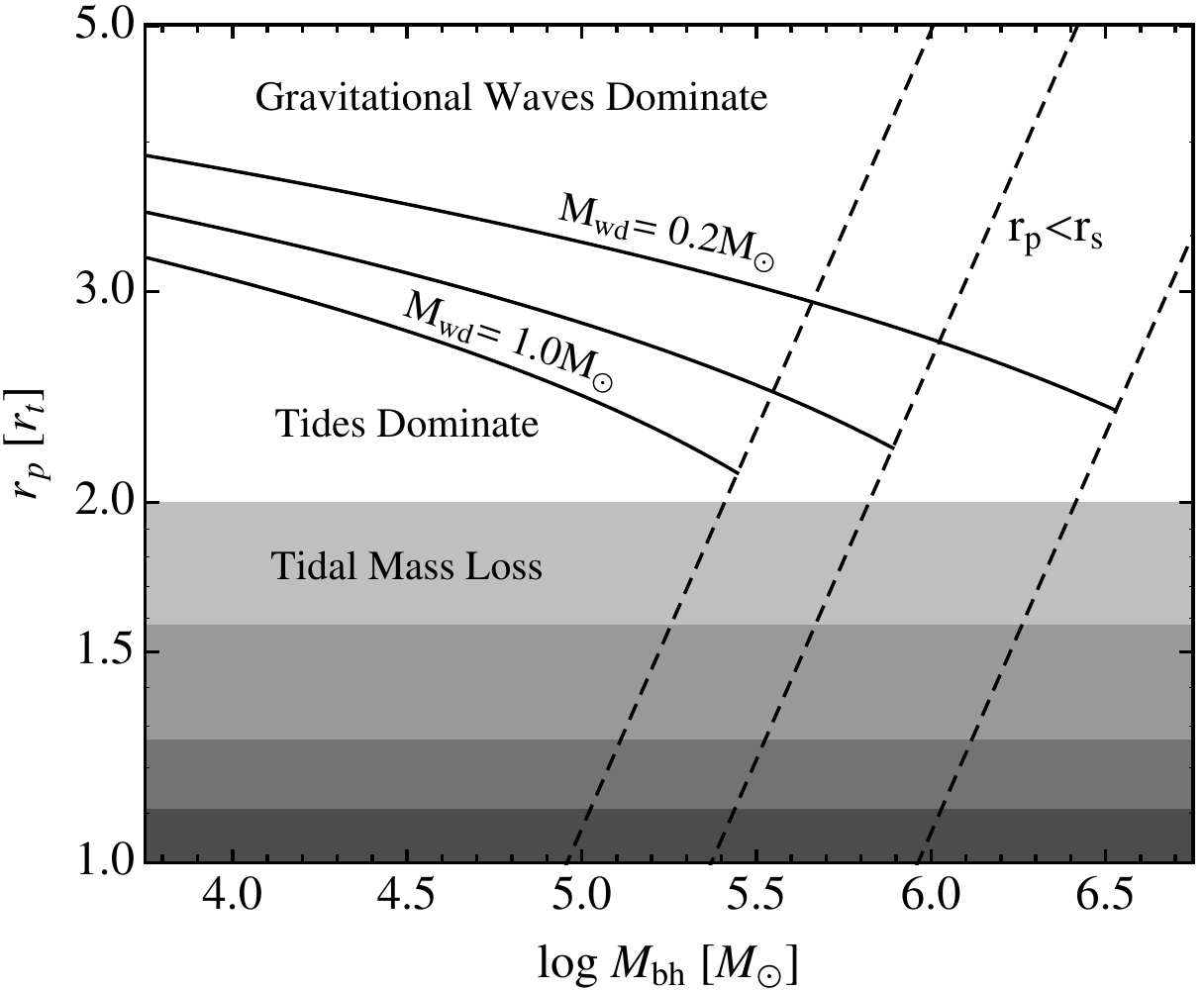}
\caption{ Phase space of encounters between WDs and MBHs. Gravitational waves are the dominant orbital evolution term above the solid lines (shown for $M_{\rm wd} = 0.2$, $0.6$, and $1.0M_\odot$). Tidal excitation is the dominant orbital-energy loss term for pericenter distances below the solid line. To the right of the dashed lines, the pericenter distance is within the MBH's Schwarzschild radius and the WD would be swallowed whole. The gray shaded area (valid to the left of the dashed lines) shows the region in which tidal forcing at pericenter is strong enough to produce mass loss. Progressive shadings show the onset of mass loss, and $\Delta M $ = 10\%, 50\% and 100\% of the WD mass, from top to bottom, respectively.  }
\label{fig:terms}
\end{center}
\end{figure}

 Another criteria we place on WD-capture orbits is that their gravitational radiation inspiral time be less than their isolation time. A chance close encounter is possible, but more likely is that another star is captured into a similarly tightly bound orbit. This unstable configuration can persist only as long as the stellar orbits avoid intersection. Eventually, this out-of-equilibrium configuration is destroyed, and one or both stars are scattered to more loosely bound orbits (or perhaps even a tidal disruption).  Thus, we take the isolation time for a captured WD to be  the inverse of the rate at which new binaries are split and deposit WDs into orbits with $a<r_{\bar m_*}$. This is, of course, an approximation and Nbody simulations offer the possibility to determine the time between exchanges of most-bound cluster stars -- with the effects of mass segregation almost certainly playing a role \citep{Gill:2008cv}. 
We therefore make a final cut that requires $t_{\rm insp} < t_{\rm iso}$, where $t_{\rm iso}$ is the isolation time as described above, and $t_{\rm insp}$ is approximated as
\beq
t_{\rm insp} \approx {a \over \dot a} \approx \frac{ c^5 a^4 (1-e^2)^{7/2}}{ G^3 \Mbh M_{\rm wd} (\Mbh + M_{\rm wd}) },
\eeq 
the order of magnitude gravitational wave inspiral time \citep{Peters:1964bc}. Gravitational radiation is the relevant loss term (as opposed to, for example, tides) because the orbits limited by this criteria are in the gravitational wave dominated regime of pericenter distance (Figure \ref{fig:terms}). 

The combination of these limits on the captured WD population ensures that these WDs will interact primarily with the MBH over the course of their orbital inspiral. In the next subsections, we describe how interactions with the MBH transform the captured distribution.

\subsection{Modeling the Evolution of Captured WD orbits}

To model the subsequent evolution of the WD orbits under the influence of both tides and gravitational radiation, we have developed an orbit-averaged code that can rapidly trace these inspirals. 
The effects of gravitational radiation from the orbit are applied following the prescription of \citep{Peters:1964bc}, which is equivalent to the 2.5-order post-Newtonian approximation. Tidal excitation is computed following the model of \citet{Mardling:1995hx,Mardling:1995it}.  \citet{Mardling:1995hx} shows that the exchange between orbital and oscillation energy depends on the amplitude and phase of the WD's oscillation as it encounters the MBH. This process leads to a ``memory" of previous interactions, and orbits that evolve chaotically as a given interaction can lead to either a positive or negative change in orbital energy and angular momentum.  To model the fiducial change in orbital energy (for an unperturbed star) we follow the prescription given by \citet{Ivanov:2007fh}
\beq\label{dE}
\Delta E_{\rm t} = 0.7 \phi^{-1} G m_{\rm wd}^2/r_{\rm wd}, 
\eeq
where $\phi = \eta^{-1} \exp\left( 2.74(\eta-1)\right) $, with the dimensionless variable $\eta$ a parameterization of pericenter distance $\eta^2 = \left(\rp/r_{\rm wd}\right)^3  \left( m_{\rm wd}/\Mbh \right)$. This expression is a fit to results computed following the method of \citet{1977ApJ...213..183P} and \citet{Lee1986}, where the overlap of $l=2$ fundamental oscillation mode with the tidal forcing is integrated along the orbital trajectory. We compared Equation \eqref{dE} with numerical results derived computing such an integral and found at most a few percent difference as a function of $\rp$, and thus we adopt this simplifying form. 
The orbital energy lost through tides goes into the quadrupole fundamental mode of the WD, which oscillates with an eigenfrequency $\omega_f \approx 1.445  G M_{\rm wd}   R_{\rm wd}^{-3}$ \citep{Ivanov:2007fh}.
The angular momentum exchange with oscillations is related to the energy loss,
\beq
\Delta L_{\rm t } = 2 \Delta E_{\rm t}/\omega_{ f}.
\eeq
Finally, we allow gravitational radiation to carry away oscillation energy from the tidally-excited WD. The luminosity of gravitational radiation scales with the oscillation energy \citep{Wheeler:1966jp}, resulting in a constant decay time of 
\beq
t_{\rm dec} = 1.5 \times 10^2 \text{yr} \left( M_{\rm wd} / M_\odot \right)^{-3} \left( R_{\rm wd} /10^{-2} R_\odot \right)^4,
\eeq
which corresponds to $t_{\rm dec} = 6447\text{ yr}$ for the $0.5M_\odot$ WD example used here \citep{Ivanov:2007fh}.  

We terminate the evolution when one of several criteria are reached:
\begin{enumerate}
\item The pericenter distance is less than the radius at which mass loss occurs $(\rp < 2 \rt)$. 
\item The accumulated oscillation energy of the WD exceeds its binding energy, 
\beq
E_{\rm osc} > {3 \over 10-2n} {G M_{\rm WD}^2 \over R_{\rm WD}}
\eeq
with $n=3/2$.  
\item The orbit circularizes. In the code this is when $e<0.1$. Further evolution is traceable via the gravitational wave inspiral as impulsive excitation of tidal oscillations no longer occurs. 
\end{enumerate}
These termination criteria correspond roughly to the categories of interactions between WDs and MBHs outlined in Section \ref{sec:sig}. When criteria 1 is met, either a single-passage tidal disruption or multiple passage mass transfer episode can be initiated depending on the orbital eccentricity, $e$.  When criteria 2 is met, eccentric-orbit mass transfer ensues. When criteria 3 is met, the WD's orbit evolves to eventual Roche-lobe overflow. 
In the next subsection, we use these termination criteria to examine the distribution of orbits at the onset of mass transfer -- after they have been transformed by their tidal and gravitational wave driven inspiral.

\subsection{Distributions at the Onset of Mass Transfer}

\begin{figure}[tbp]
\begin{center}
\includegraphics[width=0.44\textwidth]{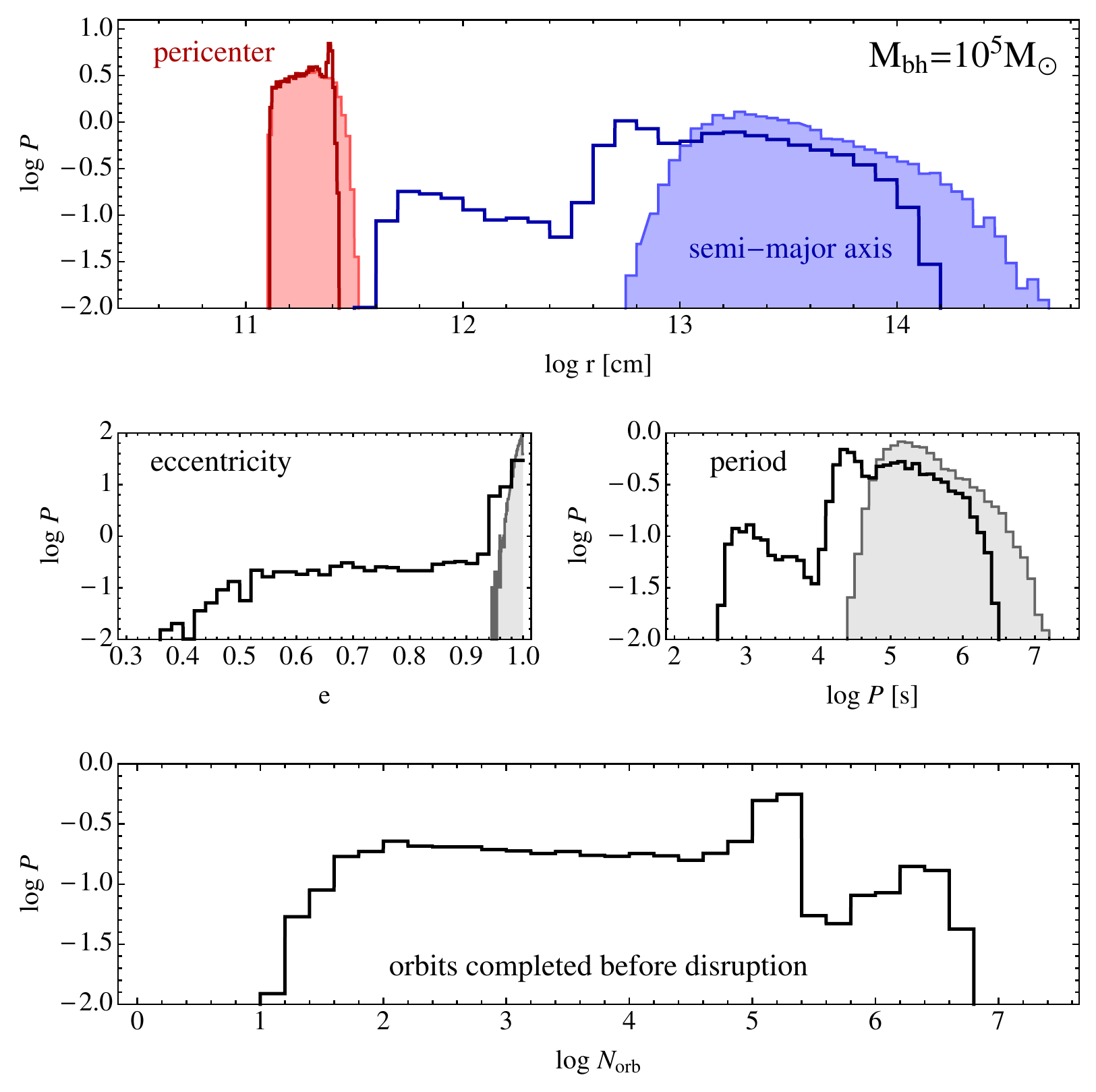}
\caption{Orbital distributions of WDs captured from split binaries at the onset of mass transfer to the MBH. Initial distributions are shown filled, final distributions are shown as lines. The upper panel shows the semi-major axis (blue) and pericenter distance (red) along with their corresponding initial distributions. The middle panels show the corresponding eccentricity and orbital period distributions. Orbits are evolved under the influence of gravitational waves and tidal excitation until the $l=2$ oscillation energy grows to reach the WD binding energy, at which point mass will be stripped from the WD envelope. The lower panel shows the number of orbits the WD survives before the onset of mass transfer, $N_{\rm orb}$.   }
\label{fig:orbdist}
\end{center}
\end{figure}

In Figure \ref{fig:orbdist}, we show how captured $0.5 M_\odot$ WDs from split binaries are eventually disrupted by a $10^5 M_\odot$ MBH.  The distribution of captured orbits is shown filled, while the final distribution is shown with lines. We find that in all cases, the deposition of orbital energy into tidal oscillation energy of the WD eventually reaches and exceeds the WD binding energy. We terminate our calculations at this point  (termination criteria 2) as this represents the onset of tidally-forced mass loss from the WD to the MBH \citep{Baumgardt:2006fg}. We find no cases of complete circularization in our simulations  (termination criteria 3). Circularization without tidal disruption requires larger initial pericenter distances, where tidal excitation is minimal, and correspondingly longer isolation times in order to allow the gravitational wave inspiral to complete \citep[e.g.][]{2004ApJ...616..221G}. The circularization and inspiral times are similar under the influence of gravitational radiation. As a result, we find no cases of termination criteria 1 in our isolated evolutions. Instead, when the pericenter distance drops to within the radius at which tides are the dominant $\Delta E$ (Figure \ref{fig:terms}), the tidal oscillation energy tends to rapidly grow to exceed the WD's binding energy leading termination criteria 2 to be met.  

The number of orbits elapsed  after capture and before the onset of mass transfer and termination is shown in the lower panel of Figure \ref{fig:orbdist}. Following the onset of mass loss, tidal stripping and eventual disruption over repeated pericenter passages proceeds as described in Section \ref{sec:sig}. 
We find that most WDs are disrupted with moderate eccentricity and a broad range of orbital periods between $10^3$ and $10^6$ s.
 The eccentricity distribution shows no nearly-circular orbits, but many orbits with $e<e_{\rm crit}$, equation \eqref{ecrit}.
 The orbital period is particularly important in the case of eccentric encounters because it sets the timescale for repetition between subsequent pericenter mass-stripping episodes. After the onset of mass transfer, the WD can be expected to survive for at most tens of passages (Section \ref{sec:sig}), thus the repetition time also fixes the range of possible total  event durations.

\section{Detecting High Energy Signatures of White Dwarf Disruption}\label{sec:detection}
In this Section, we compare the relative rates and expected luminosities of different classes of transients associated with WD-MBH interactions to discuss their detectability. We show that although rare, WD transients should outnumber their main-sequence counterparts in high energy detections because of their substantially higher peak luminosities. We then calculate that the rate of these events is sufficiently high to allow their detection by instruments such as {\it Swift}.

\begin{figure}[tbp]
\begin{center}
\includegraphics[width=0.42\textwidth]{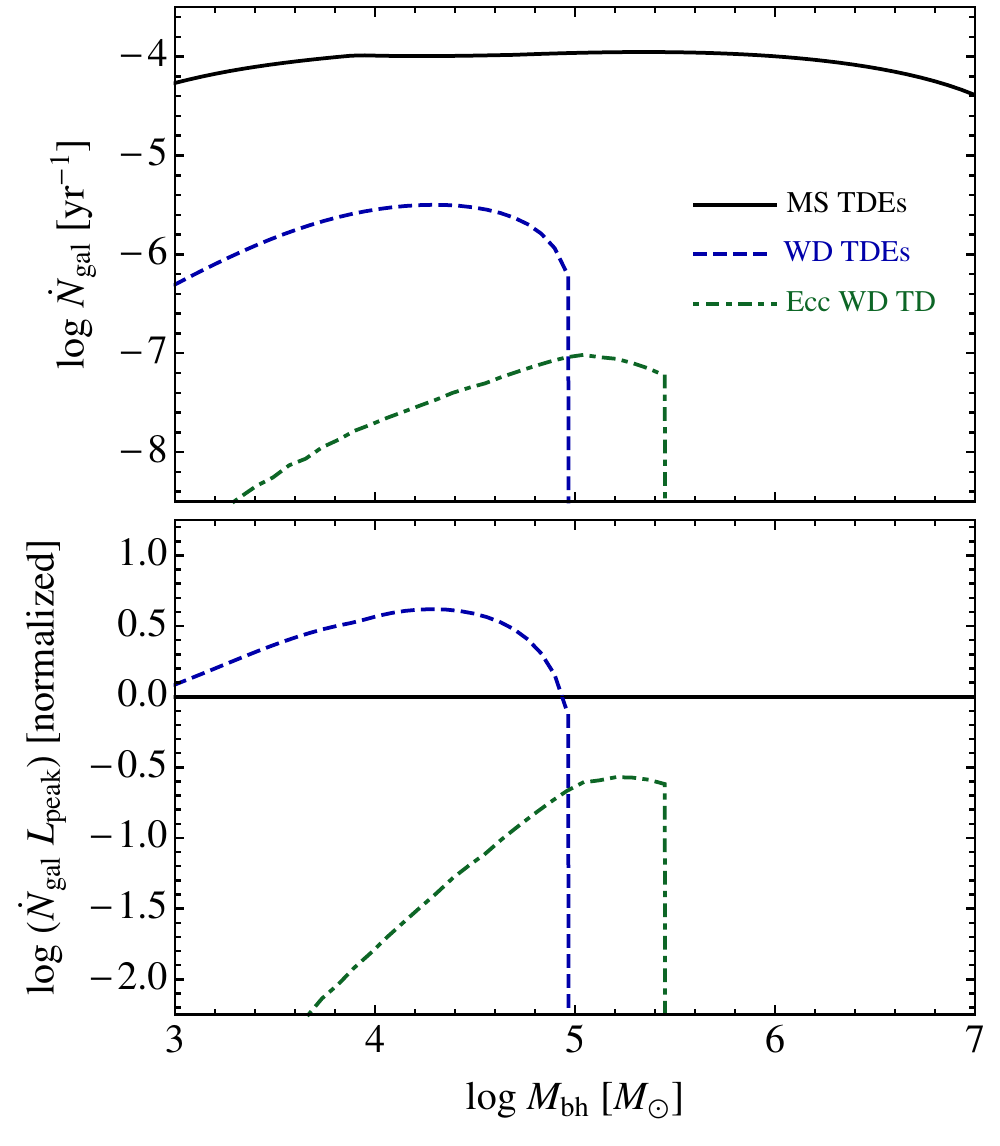}
\caption{Rates of different interaction channels per galaxy, $\dot N_{\rm gal}$, as a function of $\Mbh$. The black line is the disruption of sun-like stars. Blue is the disruption of WDs, Green is the capture of WDs by split binaries into inspiralling orbits. {\it Top:} The disruption of MS stars per galactic center greatly outnumbers that of WDs. WD disruptions peak at lower $\Mbh$ and are consumed whole by MBHs with masses $\Mbh \gtrsim 10^5 M_\odot$. Repeating flares extend to slightly higher $\Mbh$ because they are disrupted progressively with pericenter distances moderately outside the tidal radius. {\it Bottom:} When weighted by their relative luminosities, disruptions of WDs appear more common than disruptions of MS stars. This panel is normalized to the MS value, and assumes similar $f_{\rm beam}$ for all classes of events. Repeating flares are also quite luminous, but their relative rarity implies that they should make only a fractional contribution to the population of relativistic MS disruptions.  }
\label{fig:rates}
\end{center}
\end{figure}

Main sequence disruptions significantly outnumber WD disruptions and mass transfer interactions.
In the upper panel of Figure \ref{fig:rates}, we compare the rate of main-sequence star tidal disruptions to that of WD tidal disruptions, and to repeating flares resulting from mass transfer from captured WDs. The disruption rates of stars and binaries are computed by integrating the flux into the loss cone given the cluster properties outlined in \ref{sec:rates}, equation \eqref{lcrate}. In the case of repeating transients, our disruption rate calculation is supplemented by the Monte Carlo simulation that traces orbits to the onset of mass transfer, described in \ref{sec:inspiral}. To compute the values shown in the Figure, we assume a binary fraction of $f_{\rm bin}=0.1$, and a WD fraction of $f_{\rm wd}=0.1$ that applies both within the cluster and within binaries. We represent the remaining stars as main sequence stars that are sun-like, with $R_* = R_\odot$ and $M_*=M_\odot$.  

White dwarf interactions display a cut-off in MBH mass where most events transition from producing flares (if $\rp \gtrsim r_{\rm ISCO} \approx 4 \rs$) to being consumption events with little or no electromagnetic signature. For the $0.5 M_\odot$ WDs plotted, this cutoff occurs at black hole masses very near $10^5 M_\odot$. Interestingly, the progressive disruption of WDs in eccentric orbits extends to slightly higher MBH masses, since the WD is disrupted gradually, over a number of orbits, without actually penetrating all the way to the tidal radius. 
These limits in black hole mass are flexible depending on the spin parameter and orientation of the MBH's spin, since the general relativistic geodesic deviates substantially from a Newtonian trajectory in such deeply-penetrating encounters \citep{Kesden:2012cn}.  If oriented correctly with respect to a maximally rotating Kerr hole, a $0.5 M_\odot$ WD could, marginally, be disrupted by a $10^6 M_\odot$ black hole. A realistic spectrum of WD masses would also contribute to softening this transition from flaring to consumption. While the lowest mass WDs are expected to be rare in nuclear clusters due to the effects of mass segregation \citep[e.g.][]{Alexander:2005ij}, they are less dense than their more massive counterparts and could be disrupted by slightly more massive black holes. For example, a $0.1 M_\odot$ WD could be disrupted by a $3 \times 10^5 M_\odot$ black hole.  

Although rare, relativistic WD transients significantly outshine their main sequence counterparts \citep{RamirezRuiz:2009gw}. 
In the lower panel of Figure \ref{fig:rates}, we combine the relative rates of different tidal interactions with their expected peak luminosities as a function of MBH mass. We allow the beamed luminosity of all of these jetted transients to trace the mass supply to the black hole, $L\propto \dot M c^2$, as in Figure \ref{fig1} and assume that the degree of collimation is similar for each of the different classes of events. Given a population of MBHs with masses  $\Mbh\lesssim 10^5 M_\odot$, WD tidal disruptions should be more easily detected than main sequence disruptions. Eccentric disruptions over the course of multiple orbits favor slightly higher black hole masses. Their rarity compared to single-passage WD tidal disruptions implies that although they have similar peak luminosities they represent a fractional contribution to the range of detectible events.  This result suggests that WD disruptions, rather than main sequence disruptions, should serve as the most telling signpost to MBHs with masses less than $10^5 M_\odot$. In the following subsection, we discuss how high energy emission can be produced in these transients. 

\subsection{Dissipation and Emission Mechanisms}
Internal dissipation leading to a non-thermal spectrum, to be most effective, must occur when the jet is optically thin.  Otherwise it will suffer
adiabatic cooling before escaping, and could be thermalized \citep[e.g.][]{RamirezRuiz:2005dy}. 
The comoving density in the jet  propagating with a  Lorentz factor $\Gamma$ is $n'\approx L_{\rm j}/(4\pi r^2 m_p c^3 \Gamma^2)$, and using  the definition of the Thomson optical depth in a continuous outflow $\tau_{\rm j}\approx n'\sigma_{\rm T} (r/\Gamma)$ we find the location of  the  photosphere 
\begin{equation}
r_\tau={\dot{M} \sigma_{\rm T} \over 4\pi m_p c \Gamma^2}=  10^{13}\left({L_{\rm j} \over 10^{49}\;{\rm erg/s}}\right)\left({\Gamma \over 10}\right)^{-3}\;{\rm cm}.
\end{equation}

If the value of $\Gamma$ at the jet base increases by at least a  factor 2 over a timescale $\delta t$, then the later ejecta will catch up \citep{DeColle:2012bq} and dissipate a significant fraction of their kinetic energy at some distance given by \citep{Rees:1994hy}
\begin{equation}
r_\iota\approx c \delta t \Gamma^2= 3 \times 10^{13} \left({\delta t \over 10\;{\rm  s}} \right)   \left({\Gamma \over 10} \right)^2\;{\rm cm}.  
\end{equation}
Outside $r_\tau$, where radiation has decoupled from the plasma, the relativistic internal motions in the comoving frame will lead to shocks in the gas \citep{DeColle:2012bq}. 
This implies the following lower limit on 
\begin{equation}
\Gamma \gtrsim \Gamma_{\rm c}= 7.5  \left({L_{\rm j} \over 10^{49}\;{\rm erg/s}}\right)^{1/5}\left({\delta t \over 10\;{\rm s}}\right)^{-1/5}. 
\end{equation}

When $\Gamma \leq \Gamma_{\rm c}$,  the dissipation occurs
when the outflow  is optically thick and  an almost thermal transient is expected to emanate from the jet's photosphere \citep[e.g.][]{Goodman:1986hy}.  When $\Gamma \geq \Gamma_{\rm c}$, dissipation takes place when the jet is optically thin. In the presence of turbulent magnetic fields built up behind the internal shocks, the accelerated electrons within this region can  produce a synchrotron power-law radiation spectrum similar to that observed in GRBs  \citep{RamirezRuiz:2002ft,Pilla:1998eb,Meszaros:2002ks}.  
The resulting non-thermal  flare from an internal shock collision will arrive at a detector at a time $\Delta t_{\rm obs} \approx r_\iota /(c \Gamma^2) \approx \delta t$ \citep{Rees:1994hy}.
 Thus, the relative time of flare variability  at the detector will have a close one-to-one relationship with the time variability within the jet. 
 
Alternatively, high-energy emission can be produced as the jet propagates through the accretion disk region while interacting with very dense soft photon emission  with typical energy $\Theta_{\rm disk}=k T_{\rm disk} /(m_e c^2)$. A fraction $\approx \min(1,\tau_{\rm j})$ of the photons are scattered by the inverse Compton effect to energies $\approx 2\Gamma^2\Theta_{\rm disk}$, where we have assumed that a constant $\Gamma$ has been attained.  Each seed photon is boosted by $\approx \Gamma^2$ in frequency, yielding a boosted accretion disk spectrum  \citep{Bloom:2011er}.  The observed variability time scale, in this case,  is primarily related to changes in the accretion disk luminosity \citep{DeColle:2012bq}. Due to relativistic aberration, the scattered photons propagate in a narrow $1/\Gamma$ beam. The Compton drag process can be very efficient in extracting energy from the jet  and can limit its maximum speed of expansion so that $\Gamma^2 L_{\rm Edd} \lesssim L_{\rm j}$ \citep{1982MNRAS.198.1109P,RamirezRuiz:2004bi}. Typical bulk Lorentz factors range from $\Gamma \approx 10$ in quasars \citep{Begelman:1984fm}
to $\Gamma > 10^2$ in GRBs \citep{Lithwick:2001cp,Gehrels:2009fw}.
Transients that have so far been associated with tidal disruptions of stars have been mildly relativistic, with typical Lorentz factors of a few. In the case of {\it Swift} J1644+57, \citet{Zauderer:2011ie} and \citet{Berger:2012fw} inferred $\Gamma \approx 2.2$.  \citet{Cenko:2012if} find the that $\Gamma \gtrsim 2.1$ is required in {\it Swift} J2058+05. In both cases, the observed spectrum  can be explain by both internal dissipation and Compton drag \citep[see e.g.][]{Bloom:2011er}.\\
 
 \subsection{Event Rates}
 
We can estimate the detectable event rate by considering the space density of dwarf galaxies that might host these black holes. We estimate that a lower limit on the number density of dwarf galaxies is $\sim 10^7 \text{ Gpc}^{-3}$ \citep{Shcherbakov:2013hf} although recent work has shown that it may be up to a factor of $\sim30$ higher \citep{Blanton:2005gz}. If we assume that the MBH occupation fraction of these galaxies is $f_{\rm MBH}$, and adopt a per MBH rate of $\dot N_{\rm gal} \sim 10^{-6} \text{ yr}^{-1}$, then the rate of WD tidal disruptions per volume is $\dot N_{\rm vol} \sim 10 f_{\rm MBH} \text{ Gpc}^{-3} \text{ yr}^{-1}$. Note that this rate is approximately a factor of 100 smaller than the rate estimate of \citet{Shcherbakov:2013hf}, because they adopt a higher $\dot N_{\rm gal}$ that is derived by combining the tidal disruption rate normalization of an isothermal sphere $(\nu_* \propto r^{-2})$  \citep{Wang:2004jy} with the fraction of disrupted WDs from N-body simulations of globular clusters \citep{Baumgardt:2004jf,Baumgardt:2004dx}. 

Considering their high luminosity, these transients may be detected out to cosmological distances. As an example, the annual event rate for transients with $z<1$ is 
$
\dot N_{z<1} \sim 1500 f_{\rm MBH}  \text{ yr}^{-1},
$
where we have used the fact that in an $\Omega_{\rm m}=0.3$, $H_0 = 70$ cosmology, $z<1$ encloses a comoving volume of approximately 150 Gpc$^3$ \citep{Wright:2006he}. Because the emission is beamed, only a fraction $f_{\rm beam}$ are detectable from our perspective due to the random orientation of the jet column. Thus we arrive at a potentially observable event rate of 
\beq
\dot N_{z<1, \rm obs} \sim 1500 f_{\rm beam} f_{\rm MBH}  \text{ yr}^{-1}. 
\eeq
If $f_{\rm beam} = 0.1$, then of order $150  f_{\rm MBH} $ events are theoretically detectable per year. The fraction of these that would have triggered {\it Swift} in the past is still not completely understood. From Figure \ref{fig2}, typical peak timescales are thousands of seconds. \citet{Levan:2014iz} suggest that <10\% of exposures have sufficiently long-duration trigger applied to detect a longer event duration event like a WD-MBH interaction  \citep[see ][for another discussion of the detection of events in this duration range]{Zhang:2014ci}. Assuming that  10\%  of the theoretically observable events are found $\left(f_{\rm Swift}=0.1\right)$, that leaves a {\it Swift} rate of $\dot N_{\rm Swift} \sim 15 f_{\rm MBH}  \text{ yr}^{-1}$. This rate is low compared to the typical GRB rate detected by {\it Swift}, but potentially high enough to build a sample of events over a several year observing window with some long-cadence observations tailored to trigger on transients of this duration.

\section{Discussion}\label{sec:discussion}

\subsection{The MBH mass function}

For MBH masses of $\lesssim 10^5 M_\odot$, jetted transients associated with WD tidal disruptions are extremely luminous and fuel the black hole above the Eddington limit for nearly a year. These events offer a promising observational signature of quiescent black holes in this lower mass range due to their high luminosities. 
Unbeamed emission from the accretion flow is roughly Eddington-limited \citep[e.g.][]{Guillochon:2014in},
and therefore will be at least three orders of magnitude fainter than the beamed emission \citep{Haas:2012ci,Shcherbakov:2013hf}. 
While previous {\it Swift} trigger criteria catered to much shorter-duration events \citep{Lien:2014fa}, with increasing focus on long duration events recently \citep[e.g.][]{Levan:2011hq,Cenko:2012if,Levan:2014iz}, the fraction of transients that would trigger {\it Swift}, $f_{\rm Swift}$, is likely to increase or at least become better constrained in future observations. 

With a {\it Swift} detection rate of order of $\dot N_{\rm Swift} \sim 15 f_{\rm MBH} \left(f_{\rm Swift}/0.1\right) \left(f_{\rm beam}/0.1\right) \text{ yr}^{-1}$, it should be possible to constrain the occupation fraction, $f_{\rm MBH}$. More than one event per year would result if $f_{\rm MBH}\gtrsim0.1$, and thus it is most likely possible to constrain  $f_{\rm MBH}$ to that level or larger. 
In placing such a limit, there remains some degeneracy, for example, $f_{\rm MBH}= 0.1$ in the above expression could either mean that 10\% of dense nuclei harbor MBHs, or that 10\% of MBHs are surrounded by stellar systems.  
Even so, with knowledge of the expected signatures, the detection or non-detection of WD-disruption transients can place interesting constraints on the population of MBHs in this mass range with current facilities. Non-detections of events, therefore, would argue against the presence of MBHs or the presence of stellar cusps for this mass range.

\subsection{Ultra-long GRBs as WD Tidal Disruptions?}

There is tantalizing evidence that tidal disruptions of WDs by MBHs have already been detected, under the guise of ultra-long GRBs \citep{Shcherbakov:2013hf,Jonker:2013es,Levan:2014iz}.   
\citet{Levan:2014iz} elaborate on the properties of several members of the newly emerging class of ultra-long GRBs: GRB 101225A, GRB 111209A, and GRB 121027A. All of these GRBs reach peak X-ray luminosities of $\sim 10^{49} \text{erg s}^{-1}$ and non-thermal spectra reminiscent of relativistically beamed emission. At times greater than $10^4$ seconds all of these bursts exhibit luminosities that are more than a factor of a hundred higher than typical long GRBs. Astrometrically, the two bursts for which data is available (GRB 101225A are GRB 111209A) are coincident with their host galaxy's nuclear regions, suggesting compatibility with the idea that these transients originated through interaction with a central MBH. However, it is worth noting that if these events are associated with dwarf or satellite galaxies, they might appear offset from a more luminous central galaxy despite being coincident with the central regions of a fainter host, a clear-cut example being the transient source HLX-1 \citep{Farrell:2009gt}. 
\citet{Jonker:2013es} discuss a long-duration x-ray transient, XRT 000519, with a faint optical counterpart and quasi-periodic precursor emission.  The source is located near M86.  If it is at the distance of M86, the luminosity is similar to the Eddington limit of a $10^4M_\odot$ MBH. If it is, instead, a background object, the emission could be beamed and have a luminosity of up to $\sim 10^{48}$ erg s$^{-1}$.  

\begin{figure}[tb]
\begin{center}
\includegraphics[width=0.44\textwidth]{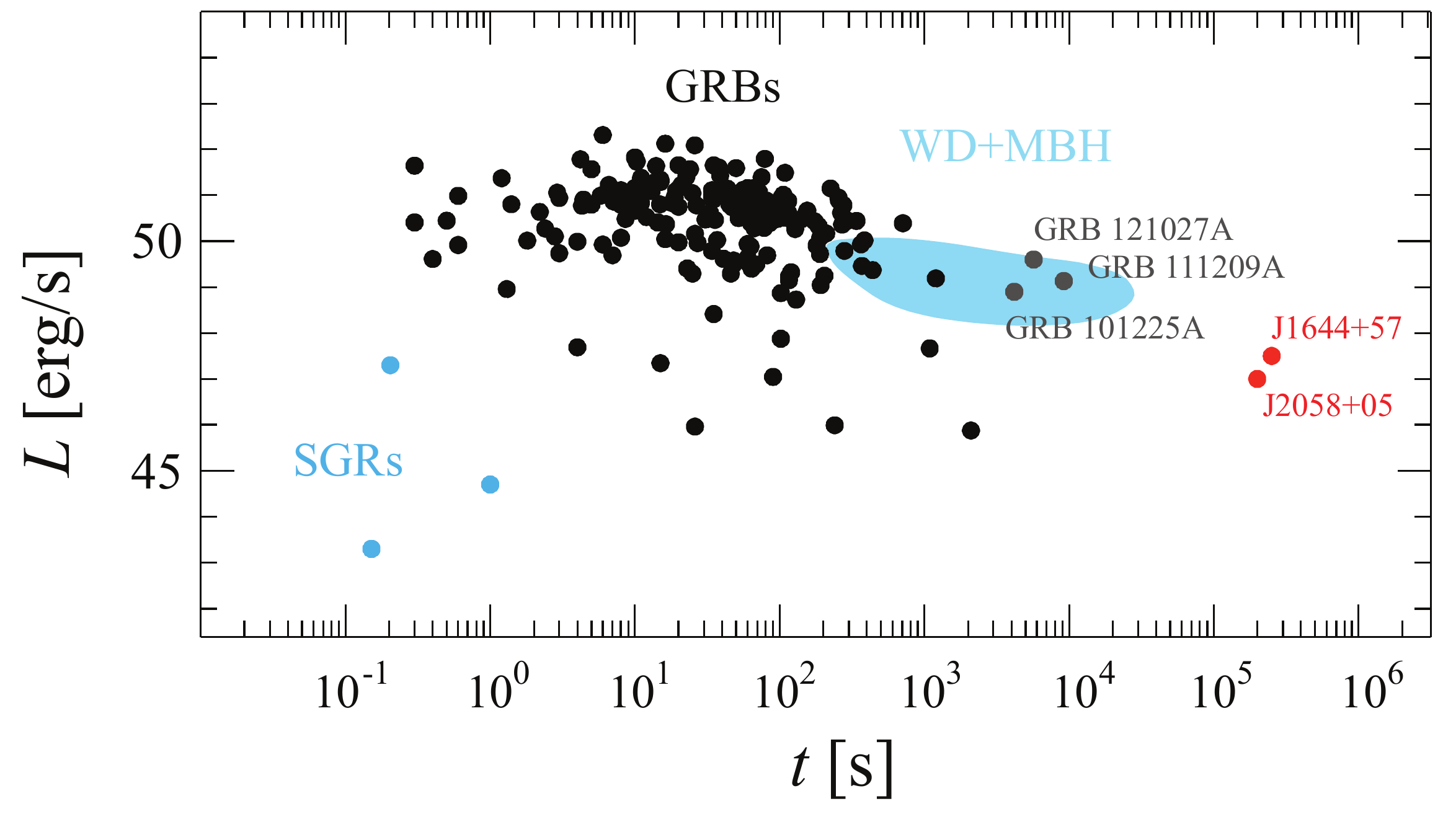}
\caption{Luminosity  versus duration adapted from \citet{Levan:2014iz}. The WD+MBH region is the region of peak timescale and luminosity for a range of WD-MBH single-passage disruptive encounters. In the shaded region, MBH masses range from $10^3$ to $10^5 M_\odot$, while the WD masses plotted are 0.25-1$M_\odot$. For the GRB and SGR sources, $t_{90}$ is plotted. If $L\propto \dot M$ in the WD disruptions, $t_{90}$ is a factor $\approx 30$ greater than $t_{\rm peak}$.  The timescales and durations of WD-MBH interactions are well removed from typical long GRBs, but coincide with those of the emerging class of ultra-long GRBs, such as GRB 101225A, GRB 111209A, and GRB 121027A.}
\label{fig:phasespace}
\end{center}
\end{figure}

Might such events be tidal disruptions of WDs by MBHs? 
Further evidence is certainly needed to ascertain the origin of these bursts, but the properties, including luminosities and decay timescales are in line with those we have reviewed for disruptions of WDs by MBHs. 
Figure \ref{fig:phasespace} augments the phase space diagram of \citet{Levan:2014iz}, showing characteristic luminosities and decay times for single-passage tidal disruptions of WDs and MBHs (blue shaded region). In Figure \ref{fig:phasespace}, we plot the peak timescale and luminosity of peak for the disruptions, for MBH masses from $10^3$ to $10^5M_\odot$, and WD masses of 0.25 - 1$ M_\odot$. Other relevant timescales include $t_{\rm Edd}$, the time above the MBH's Eddington limit, plotted in Figure \ref{fig2}, and $t_{90}$, as plotted for the GRB and soft gamma-ray repeater (SGR) sources, which is a factor of  $\approx 30$ greater than $t_{\rm peak}$.

The peaks in the lightcurve of {\it Swift} J1644+57 \citep[e.g.][]{Saxton:2012ip} have been associated with periodic spikes in the mass supply from a gradually disrupting WD in an eccentric orbit by \citet{Krolik:2011ew}.  The suggested repetition time is $P\sim5\times10^4$s  \citep{Krolik:2011ew}.  In our $\Mbh =10^5 M_\odot$, $M_{\rm wd} = 0.5M_\odot$ example of Figure \ref{fig:orbdist}, $\sim40$\% of the captured population initiates mass transfer with orbital periods $10^4{\rm s}<P<10^5{\rm s}$, thus, reproducing this repetition time does seem to be possible.  
Our inspiral simulations suggest that such repeating encounters are approximately an order of magnitude less common than their single-passage WD-dispruption counterparts. 
More importantly for determining the origin of {\it Swift} J1644+57, by comparison to Figure \ref{fig:rates} we expect that repeating encounters with these sorts of repetition times would be detected at $\sim 10$\% the rate of jetted main-sequence disruptions from these same MBH masses. 
However, single-passage WD disruptions, repeating encounters, and main-sequence disruptions each originate from different range of characteristic MBH masses (as shown in the lower panel of Figure \ref{fig:rates}). If there is a strong cutoff in the low end of the MBH mass function we might expect this to truncate one class of events but not another. 

One remaining mystery is the shape of the lightcurve of {\it Swift} J1644+57 during the plateau phase. Variability could originate in modulated mass transfer \citep{Krolik:2011ew} or from the accretion flow and jet column itself, as described in Section \ref{sec:detection}, \citep[and by][]{DeColle:2012bq}. If the jetted luminosity traces the mass accretion rate, $L \propto \dot M c^2$, as we have assumed here, we would expect  the peaks in {\it Swift} J1644+57's lightcurve to trace the exponentiating mass loss from the WD -- instead of the observed plateau. If, however, this simplifying assumption proves incorrect (or incomplete) it does appear to be possible to produce events with plateau and super-Eddington timescales comparable to {\it Swift} J1644+57 with multi-passage disruptions of WDs. Detailed simulations of disk-assembly in multi-passage encounters offer perhaps the best hope to further constrain the electromagnetic signatures of these events.

In WD disruptions, the jetted component is significantly more luminous than the Eddington-limited accretion disk component \citep[about a thousand times more-so than in the main sequence case;][]{DeColle:2012bq,Guillochon:2013jj}, and thus we have pursued the beamed high-energy signatures of these events in this paper. 
With the advent of LSST, however, detecting the corresponding  disk emission signatures may become more promising.  In a fraction of events that  pass well within the tidal radius \citep[e.g.][]{Carter:1982fn,Guillochon:2009di}, a detonation might be  ignited upon compression of the WD  \citep{Luminet:1989wl,Rosswog:2009gg,Haas:2012ci,Shcherbakov:2013hf}.  In this scenario, maximum tidal compression can cause the shocked white dwarf material to exceed the threshold for pycnonuclear reactions so that thermonuclear runaway ensues \citep{Holcomb:2013et}. The critical $\beta$ appears to be $\gtrsim 3$  \citep{Rosswog:2009gg}, so perhaps $\lesssim 1/3$ of the high-energy transients plotted in Figure \ref{fig:phasespace} are expected to be accompanied by an  optical counterpart in the form of an atypical type I supernova.

Robustly separating ultra-long GRBs into core collapse and tidal disruption alternatives remains a challange   \citep[see e.g.][]{Gendre:2013bs,Boer:2013tc,Stratta:2013er,Yu:2013vh,Levan:2014iz,Zhang:2014ci,Piro:2014wy}. The central engines of ultra-long GRBs are essentially masked by high-energy emission with largely featureless spectra, revealing little more than the basic energetics of the relativistic outflow \citep{Levan:2014iz}. Several distinguishing characteristics are, however, available. Variability timescales should be different (as they would be associated with compact objects of very different mass, see Section \ref{sec:detection}). Significantly, the evolution of the prompt and afterglow emission at high energy and at radio wavelengths would be expected to deviate from that of a canonical impulsive blast wave in tidal disruption events due to long-term energy injection from the central engine \citep{DeColle:2012bq,Zauderer:2013jr}.  Disk emission, if detected in optical or UV observations, would present strong evidence of tidal disruption origin. While the bulk of WD disruptions would lack a coincident supernova,  a minority would be accompanied by atypical type I supernovae. Optical signatures of a core-collapse event are uncertain, perhaps involving emission from the cocoon \citep{RamirezRuiz:2002hx,Kashiyama:2013cb,Nakauchi:2013kg}, accretion disk wind \citep{MacFadyen:1999kr,Pruet:2004fc,LopezCamara:2009iq}, or type IIP-like lightcurves \citep{Levan:2014iz} but the detection of hydrogen lines in an accompanying supernova spectrum would point to a core-collapse origin. \citet{Levan:2014iz} emphasize that one way to tackle these observational challenges in the near term is looking statistically at the astrometric positions of ultra-long bursts and whether they coincide with galactic centers. 

\subsection{Prospects for simultaneous Electromagnetic and Gravitational Wave Detection}

A primary source of interest  in WD-MBH interactions has been their potential as sources of both electromagnetic and gravitational wave emission 
\citep{Ivanov:2007fh,Rosswog:2008gc,Rosswog:2008hv,Sesana:2008ii,Rosswog:2009gg,Zalamea:2010eu,Haas:2012ci,Dai:2013bt,Cheng:2013cm}, especially as these events, if observed, would constrain the MBH mass function at low masses \citep[e.g.][]{deFreitasPacheco:2006hh}. Chirp waveforms have been computed for single, disruptive passages \citep[e.g.][]{Rosswog:2009gg,Haas:2012ci}  and should be detectible only if the source is within  $\sim 1$Mpc  given a $10^5 M_\odot$ MBH \citep{Rosswog:2009gg}.

Potentially less restrictive are longer-lived periodic signals \citep[though, see][]{Berry:2013fg}. 
The longest-lived transient, and that with the most uniform periodicity, would occur if a WD were overflowing its Roche lobe and transferring mass to the MBH from a circular orbit \citep[e.g.][]{Dai:2013bt}. However, we see no such circularization events in our orbit evolution simulations. 
Instead, the build-up of tidal oscillation energy in the WD leads to its disruption before the orbit circularizes, even in cases where gravitational radiation is the dominant term in the orbit evolution. 
In these eccentric cases, the gravitational wave signature reminiscent would be of a series of roughly-periodically spaced chirps associated with the pericenter passages. 
It is worth noting that these passages should not be strictly periodic because the orbital period wanders chaotically as successive passages pump energy into and out of the WD oscillations depending on the oscillation phase with which it encounters the MBH \citep{Mardling:1995hx,Mardling:1995it}.

\section{Summary}
In this paper we have discussed the role that orbital dynamics plays in shaping the transients that result from interactions between WDs and MBHs. 
WDs most commonly encounter black holes in single passages. Multiple passages from an eccentric orbit are about an order of magnitude less common, but would have characteristic repetition timescales of $10^4 - 10^6$ s. 
 The relative paucity of repeating events in our calculations, combined with the small range of MBH masses in which they appear to occur, suggests that the likelihood that {\it Swift} J1644+57 could form via the repeating disruption channel, as outlined shortly after the event by \citet{Krolik:2011ew}, is $\lesssim 10$\%.  
We find no instances of mass transfer from a circular orbit. The consequence of these encounters is a mass supply that greatly exceeds the MBH's Eddington limit. We expect the resulting thick accretion flow should amplify a poloidal magnetic field and lunch a jet. The relativistically beamed emission from these events may be more readily detectable than beamed emission from disruptions of main sequence stars. We therefore argue that the best prospects to constraining the lower-mass end of the MBH mass function lie in searching for the high-energy signatures of WD disruption events. The possibility of collecting a sample of such events in coming years with {\it Swift} appears promising \citep[e.g.][]{Shcherbakov:2013hf,Jonker:2013es,Levan:2014iz}. The detection or non-detection of these transients should offer strong constraints on the population of MBHs with masses $\Mbh \lesssim 10^5 M_\odot$ and the nature of the stellar clusters that surround them. 

\acknowledgments{
We would like to thank Fabio Antonini, Tamara Bogdanovic, Priscilla Canizares, Annalisa Celotti, Dan Kasen, Julian Krolik, Doug Lin, Ilya Mandel, Martin Rees, Stephan Rosswog, Nicholas Stone, and Michele Trenti for a variety of helpful discussions.  We also thank the anonymous referee for a thoughtful report that improved this work. We benefitted from the hospitality of the DARK cosmology centre, Institute of Astronomy at Cambridge University, and the INAF-Osservatorio Astronomico di Roma while completing this work. We acknowledge support from the David and Lucile Packard Foundation, NSF grant: AST-0847563, the NSF Graduate Research Fellowship (M. MacLeod), the NASA California Space Grant Consortium Undergraduate Research Opportunity Program (J. Goldstein) and the Einstein Fellowship (J. Guillochon). }

\bibliographystyle{apj}

\end{document}